\documentclass[12pt]{iopart}

\expandafter\let\csname equation*\endcsname\relax
\expandafter\let\csname endequation*\endcsname\relax

\usepackage{graphicx}
\usepackage{subfigure}
\usepackage{verbatim}
\usepackage{dcolumn}
\usepackage{bm}
\usepackage{color}
\usepackage[toc]{appendix}
\usepackage{stmaryrd}
\usepackage{amsthm}
\usepackage{hhline}
\usepackage{amssymb}
\usepackage{iopams}
\usepackage{cite}
\usepackage{tikz}
\usepackage{bbm}
\usepackage[mathscr]{eucal}
\usetikzlibrary{arrows,shapes,calc,matrix}
\usepackage{amsmath}
\usepackage{xstring}
\usepackage[colorlinks=true,citecolor=blue,linkcolor=blue,urlcolor=blue]{hyperref}
\usepackage{doi}

\makeatletter
\newcommand\footnoteref[1]{\protected@xdef\@thefnmark{\ref{#1}}\@footnotemark}
\makeatother

\newcommand\invisiblesection[1]{%
	\refstepcounter{section}%
	\addcontentsline{toc}{section}{\protect\numberline{\thesection}#1}%
	\sectionmark{#1}}

\newcommand{\RNum}[1]{\uppercase\expandafter{\romannumeral #1\relax}}

\usepackage{cleveref}

\begin{document}  
\title[Optically controlled thermal modulator with common reservoirs]{Heat transfer in transversely coupled qubits: Optically controlled thermal modulator with common reservoirs}

\author{Yi-jia Yang\textsuperscript{1}, Yu-qiang Liu\textsuperscript{1} and Chang-shui Yu\textsuperscript{1}}
\address{$^1$School of Physics, Dalian University of Technology, Dalian 116024, P.R. China}
\ead{ycs@dlut.edu.cn}

\vspace{10pt}
\begin{indented}
\item[]August 2022
\end{indented}

\begin{abstract}
This paper systematically studied heat transfer through two transversely coupled qubits in contact with two types of heat reservoirs. One is the independent heat reservoir which essentially interacts with only a single qubit, the other is the common heat reservoir which is allowed to simultaneously interact with two qubits.
Compared to independent heat reservoirs, common reservoirs always suppress heat current in most cases. However, the common environment could enhance heat current, if the dissipation rate corresponding to the higher eigenfrequency is significantly higher than that corresponding to the lower eigenfrequency.
In particular, in the case of resonant coupling of two qubits and the proper dissipations, the steady state can be decomposed into a stationary dark state which doesn't evolve and contributes zero heat current, and a residual steady state which corresponds to the maximal heat current. This dark state enables us to control steady-state heat current with an external control field and design a thermal modulator. In addition, we find that inverse heat currents could be present in the dissipative subchannels between the system and reservoirs, which interprets the suppression roles of common heat reservoirs. We also calculate the concurrence of assistance (COA) of the system and find that heat current and COA have the same trend with temperature, which further indicates that entanglement can be regarded as a resource to regulate heat transport.
\end{abstract}

\section{Introduction} 
\label{sec:1}
Thermodynamics, one of the pillars of physics, has been widely investigated since its birth  \cite{landsberg1956foundations,parrondo2015thermodynamics}. With the rapid development of quantum mechanics  \cite{sakurai1995modern,ballentine2014quantum}, thermodynamics at the quantum level has attracted increasing interest  \cite{zohar1990quantum,lemaitre1931beginning,ballentine2014quantum,cao2021quantum,hewgill2021quantum}. For example, Otto cycles \cite{quan2007quantum,liu2021periodically}, Carnot engines \cite{feldmann2000performance,palao2001quantum,ccakmak2020quantum}, Brownian motions \cite{humphrey2002reversible,lu2020brownian}, four laws of thermodynamics \cite{wang2002experimental,kosloff2013quantum,levy2014local}, entropy increasing principle \cite{lieb2000fresh,quan2006maxwell,strasberg2013thermodynamics,cottet2017observing} and so on are widely studied, which deepens our understanding of the thermodynamical laws down to the quantum level \cite{binder2019thermodynamics,millen2016perspective,vinjanampathy2016quantum} and provides us with the potential to constructively use quantum features to exploit quantum thermal devices that can manipulate the microscopic energy flow.

 In recent years, various quantum thermal devices based on versatile systems have been proposed such as quantum engine and refrigerator \cite{ThreeLevelMasers,yu2014re,venturelli2013minimal,hofer2016autonomous,yu2019quantum}, quantum thermometers \cite{hofer2017quantum,yang2019thermal}, thermal rectifier \cite{wang2019thermal} and switch \cite{karimi2017coupled,farsani2019quantum}, thermal diode \cite{lashkaryov1941investigations,li2004thermal,werlang2014optimal,PhysRevE.99.042121} and transistor \cite{joulain2016quantum,guo2018quantum,majland2020quantum,PhysRevResearch.2.033285,wijesekara2020optically,PhysRevA.103.052613,wijesekara2021darlington,mandarino2021thermal,e24010032} and so on. Experimental advances have also been made continuously \cite{quan2006maxwell,majer2007coupling,sillanpaa2007coherent,you2011atomic,xiang2013hybrid,koski2014experimental,pekola2015towards,hofer2016quantum,hofer2016autonomous}, thermodynamic devices at the quantum level using superconducting qubits, quantum dots, cavity QED, circuit QED, and other systems have been realized. Most cases address the individual component of a system only interacting with an independent heat reservoir (IHR) but a common heat reservoir (CHR) could produce interesting effects like increasing entanglement etc \cite{kim2002entanglement,schaller2009transport,liao2011quantum,2013Steady,hewgill2018quantum,man2019improving,cattaneo2019local}. 
In addition, two-dimensional materials have also been rapidly developed due to the various features, such as high mobility, band gap tunability, super-large specific surface area, and so on \cite{dean2010boron,li2014black,zhang2017metallic}. In particular, graphene \cite{geim2009graphene,faugeras2010thermal,novoselov2012roadmap} has good thermal conductivity and the thermal relaxation is also studied in cavity-coupled graphene with a Johnson noise read-out \cite{efetov2018fast}. In this sense, the heat current through the single layer could be greatly affected by the transverse coupling between the components (e.g., a layer of graphene molecules). 
\begin{figure}
\begin{center}
\includegraphics[width=.6\textwidth]{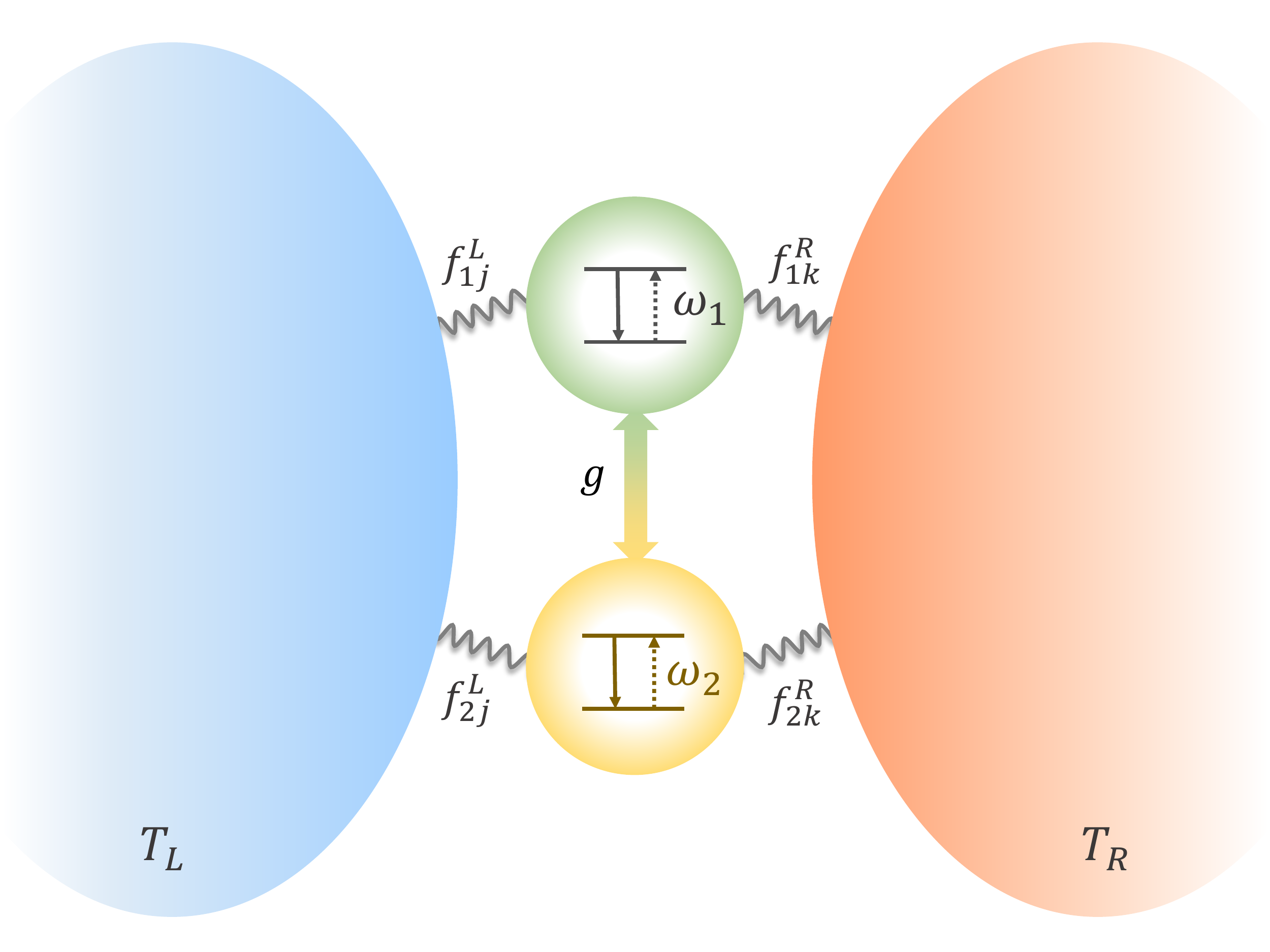}
\caption{Two coupled atoms with the transition frequencies $\omega_1 $ and $\omega_2$ are in contact with two CHRs $L$ and $R$. The coupling strength between the two TLS is $g$ and the temperatures of the two CHRs are $T_L$ and $T_R$, respectively. $f_{1j}^L$, $f_{1k}^R$, $f_{2j}^L$ and $f_{2k}^R$ denote the coupling strengths between the TLS and the corresponding modes in the CHRs.}
\label{model_common}
\end{center}
\end{figure}

In this paper, we consider a simple model consisting of two transversely coupled two-level atoms (TLAs) commonly contacting two heat reservoirs as shown in figure \ref{model_common}. We systematically consider and compare all four cases, including whether or not there exists the transverse coupling and whether or not the two TLAs are coupled with the common heat reservoirs. Using the Born-Markov-secular (BMS) master equation, we find that the transverse coupling between the two TLAs reduces the heat currents despite CHRs or IHRs. It is found that CHR can enhance or suppress the heat currents (HCs), which depends on the practical scenario such as the decay rates, the atomic coupling, the temperature, etc.
 In particular, we show that if two TLAs are resonantly coupled with each other with the same dissipation rates, the system's steady state is a mixture of two states: one is a 'dark state' that does not evolve and is decisive with zero HC; the other is a steady state leading to the maximum HC. The 'dark state' enables us to design a quantum thermal modulator that can continuously control the desired steady-state HC by an external driving laser. 
In addition, the dissipation channels between CHRs and atoms are divided into direct dissipation channels (DDCs) and cross dissipation channels (CDCs). We find that inverse currents \cite{wang2020inverse} are generated in CDCs for detuned coupling and both CDC and DDC for resonant coupling. It also gives an interpretation of the suppression effects of the CHRs.
Besides, we also study the concurrence of assistance (COA) \cite{2003Local,gour2005deterministic} in the system, which indicates that HC and COA keep a similar variation trend with the changing reservoir temperatures.

The structure of this paper is as follows. In section \ref{sec:2}, we introduce the physical model of two transversely coupled TLAs contacting two CHRs and derive the BMS master equation of the dynamic process. In section \ref{sec:3} and section \ref{sec:4}, we give the expressions of steady-state HCs with detuned coupled TLAs and resonantly coupled TLAs, respectively. In section \ref{sec:5}, we studied the 'dark state' of the system and designed a thermal modulator. In section \ref{sec:6}, we present HCs in different dissipation channels and discuss the presence of inverse currents. In section \ref{sec:7}, steady-state COA and HC are analyzed. Section \ref{sec:8} gives a summary of this paper. Some detailed derivations and tedious expressions are shown in the appendices.             

\section{The model and the master equation} 
\label{sec:2}

We consider two coupled TLAs with natural frequency $\omega_1$ and $\omega_2$  and the coupling strength $g$, which are simultaneously connected to a heat reservoir $L$ with temperature $T_L$ and a heat reservoir $R$ with temperature  $T_R$, respectively, as shown in figure \ref{model_common}.
The Hamiltonian $H$ of the system can be written as ($k_B=\hbar=1$)
\begin{equation}
H=H_S+H_E+H_{SE},
\end{equation}
where
\begin{equation}
H_S=\frac{\omega_1}{2}\sigma_1^z+\frac{\omega_2}{2}\sigma_2^z+g\sigma_1^x\sigma_2^x
\end{equation} is the Hamiltonian of the two interacting TLAs with $\sigma_{m}^z=\vert e\rangle_m\langle e\vert-\vert g\rangle_m\langle g\vert$, $m=1,2$. Here we don't apply the rotational wave approximation and hence take the X-X type interaction with $\sigma_{m}^x=\vert e\rangle_m\langle g\vert+\vert g\rangle_m\langle e\vert$. We assume that all reservoirs are composed of harmonic oscillators with infinite degrees of freedom, so the Hamiltonian $H_E$ of the environment (heat reservoirs) reads
\begin{equation}
H_E=\sum_j\omega_{Lj}a_j^\dagger a_j+\sum_k\omega_{Rk}b_k^\dagger b_k\label{H_E},
\end{equation}
where $\omega_{Lj}$ ($\omega_{Rk}$) represents the transition frequency of the $j_{th}$ mode of reservoir $L$ (the $k_{th}$ mode of reservoir $R$) and $a_j^\dagger$ and $a_j$ ($b_k^\dagger$ and $b_k$) are the corresponding creation and annihilation operators. The interaction $H_{SE}$ between the system and the reservoirs is
\begin{equation}
H_{SE}=\sum_{m=1}^2[\sum_jf_{mj}^L(a_j^\dagger+a_j)+\sum_kf_{mk}^R(b_k^\dagger+b_k)]\sigma_m^x,
\label{H_(SE)}
\end{equation}
where $f_{mj}^L$ (or $f_{mk}^R)$ denotes the coupling strength between the $m_{th}$ atom and the $j_{th}$ (or $k_{th}$) mode of thermal reservoir $L$ (or $R$), respectively.

To describe the system's dynamics, we'd like to rewrite the Hamiltonian $H$ in $H_S$ representation.
The Hamiltonian $H_S$ is a diagonalized matrix in its own representation as $H_S=\sum_{l=1}^4\lambda_l\vert l\rangle_l\langle l\vert$, where
\begin{equation}
[\lambda_1,\lambda_2,\lambda_3,\lambda_4]=[-\Gamma_s,-\Gamma_d,\Gamma_d,\Gamma_s]
\end{equation}
is the eigenvalue with $\Gamma_\nu=\sqrt{\omega_\nu^2+g^2}$, $\nu=s,d$, and $\omega_s$ or $\omega_d$ is the sum or difference between the natural frequencies of two TLA, i.e., $\omega_s=\frac{1}{2}(\omega_1+\omega_2)$ and $\omega_d=\frac{1}{2}(\omega_1-\omega_2)$, and
\begin{alignat}{2}
\vert1\rangle&=-\sin\theta_s\vert\uparrow\uparrow\rangle+\cos\theta_s\vert\downarrow\downarrow\rangle, &\quad \vert2\rangle&=-\sin\theta_d\vert\uparrow\downarrow\rangle+\cos\theta_d\vert\downarrow\uparrow\rangle,\\
\vert3\rangle&=\cos\theta_d\vert\uparrow\downarrow\rangle+\sin\theta_d\vert\downarrow\uparrow\rangle, &\quad \vert4\rangle&=\cos\theta_s\vert\uparrow\uparrow\rangle+\sin\theta_s\vert\downarrow\downarrow\rangle,
\end{alignat}
denote the corresponding eigenstates, with $\sin\theta_\nu=\frac{g}{\sqrt{(\Gamma_\nu+\omega_\nu)^2+g^2}}$, $\vert\uparrow\uparrow\rangle=\vert\uparrow\rangle_1\otimes\vert\uparrow\rangle_2$, $\vert\uparrow\rangle_m=\left(\begin{smallmatrix}1\\0\end{smallmatrix}\right)$ and $\vert\downarrow\rangle_m=\left(\begin{smallmatrix}0\\1\end{smallmatrix}\right)$. The operators $\sigma_m^x$ in $H_S$ representation is converted to the eigenoperators $V_m(\omega_{m\mu})$ as 
\begin{alignat}{2}
V_{1}(\omega_{11})&=&\sin\theta_+(\vert3\rangle\langle 4\vert-\vert1\rangle\langle 2\vert) &,\quad \omega_{11}=\omega_{-},\\
V_{1}(\omega_{12})&=&\cos\theta_+(\vert1\rangle\langle 3\vert+\vert2\rangle\langle 4\vert) &,\quad \omega_{12}=\omega_{+},\\
V_{2}(\omega_{21})&=&\cos\theta_-(\vert1\rangle\langle 2\vert+\vert3\rangle\langle 4\vert) &,\quad \omega_{21}=\omega_{-},\\
V_{2}(\omega_{22})&=&\sin\theta_-(\vert1\rangle\langle 3\vert-\vert2\rangle\langle 4\vert) &,\quad \omega_{22}=\omega_{+},
\end{alignat}
where $\theta_{\pm}=\theta_d\pm\theta_s$ and $\omega_{m\mu}, \mu=1,2$, are the eigenfrequencies with $\omega_{\pm}=\Gamma_s\pm\Gamma_d$.
It is easily checked that the eigenoperators satisfy the commutation relations $[H_S,V_m(\omega_{m\mu})]=-\omega_{m\mu}V_m(\omega_{m\mu})$ and $[H_S,V^\dagger _m(\omega_{m\mu})]=\omega_{m\mu}V^\dagger_m(\omega_{m\mu})$.
Thus the interaction Hamiltonian ${H}_{SE}$ can also be rewritten as
\begin{equation}
\tilde{H}_{SE}=\sum_{m=1}^2[\sum_jf_{mj}^L(a_j^\dagger +a_j)+\sum_kf_{mk}^R(b_k^\dagger +b_k)]\times\sum_{i=\pm}[V_{m}^\dagger (\omega_{i})+V_{m}(\omega_{i})],
\end{equation}
and the total Hamiltonian $H$ reads
\begin{equation}
H=\sum_{l=1}^4\lambda_l\vert l\rangle_l\langle l\vert+H_E+\tilde{H}_{SE}. \label {15}
\end{equation}

With the Hamiltonian equation (\ref{15}), we employ the Born-Markov-secular approximation to derive the global master equation which describes the dynamic evolution of an open system \cite{breuer2002theory}. The details of the derivation are given in appendix \hyperref[AppendixA]{A}. In the interaction picture, the master equation governing the evolution of the reduced density matrix $\rho$ of the system can be written as
\begin{equation}\label{MEq}
\dot\rho=\mathcal{L}_L(\rho)+\mathcal{L}_R(\rho),
\end{equation}
where $\mathcal{L}_{\alpha}(\rho),\alpha=L, R$ denotes the dissipators of the left or the right thermal reservoirs and
\begin{equation}
\mathcal{L}_{\alpha}(\rho)=\mathcal{L}_\alpha^{11}(\rho)+\mathcal{L}_\alpha^{22}(\rho)+\mathcal{L}_\alpha^{12}(\rho)+\mathcal{L}_\alpha^{21}(\rho)\label{MEq.(common)},
\end{equation}
with
\begin{align}
\mathcal{L}_{\alpha}^{mn}(\rho)=&\sum_{i=\pm}J_{\alpha}^{mn}(-\omega_i)[2V_{n}(\omega_i)\rho V_{m}^\dagger (\omega_i)-V_{m}^\dagger (\omega_i)V_{n}(\omega_i)\rho-\rho V_{m}^\dagger (\omega_i)V_{n}(\omega_i)\nonumber\\
&+J_{\alpha}^{mn}(\omega_i)[2V_{n}^\dagger(\omega_i)\rho V_{m}(\omega_i)-V_{m} (\omega_i)V_{n}^\dagger(\omega_i)\rho-\rho V_{m} (\omega_i)V_{n}^\dagger(\omega_i)].\label{dissipator}
\end{align}
Here $J_\alpha^{mn}(\pm\omega_i)=\gamma^{mn}_\alpha(\omega_i)[\pm\bar{n}_\alpha(\pm\omega_i)] $ is the spectral density, $\gamma_{L/R}^{mn}(\omega_i)=\pi f_{m j/k}^{L/R}(\omega_i)f_{n j/k}^{L/R}(\omega_i)$ is the dissipation rate and the average photon number of the {\it{i}}th mode for reservoir $\alpha$ is denoted by $\bar{n}_\alpha(\omega_i)=\frac{1}{e^{\frac{\omega_i}{T_\alpha}}-1}$ with temperature $T_\alpha$. It should be noted that
$\mathcal{L}_{\alpha}^{mn}(\rho)$ denotes the direct ($m=n$) or crossing ($m\neq n$) dissipations induced by the $m_{th}$ atom or two atoms with the $\alpha_{th}$ heat reservoir. In particular,
the CHR effect is indicated by the crossing dissipations. 
 If CHR is not considered, $\mathcal{L}_\alpha^{mn}(\rho)=0$, which will reduce to the case of IHRs, namely, each atom is only coupled to its own two reservoirs ($L$ and $R$), in which case we let the heat reservoirs with the same subscript ($L$ or $R$) be of the same temperature for the sake of comparison. So the corresponding master equation also can be formally given as equation (\ref{MEq}) (the detailed derivation is in appendix \hyperref[AppendixB]{B}), but the difference is 
\begin{equation}
\mathcal{L}_{\alpha}(\rho)=\mathcal{L}_\alpha^{11}(\rho)+\mathcal{L}_\alpha^{22}(\rho),\label{MEq.(independent)}
\end{equation}
which has no crossing dissipations.

In the case of CHR, if the two atoms aren't coupled to each other, the eigenoperator $V_m(\omega_{mm})=0$ due to $\sin\theta_{\pm}=0$, so only one eigenoperator is left for each atom with the corresponding eigenfrequency $\omega_{1}$, $\omega_{2}$, and the master equation is given as equations (\ref{MEq},\ref{MEq.(independent)}) for $\omega_{1}\neq\omega_{2}$ due to the secular approximation.
It is worth mentioning that there will be crossing dissipations for $\omega_{1}=\omega_{2}$, in which case the master equation is given as equations (\ref{MEq},\ref{MEq.(common)}). The explicit derivation is provided in appendix \hyperref[AppendixB]{B}.

If neither the two-atomic coupling nor the effects of CHRs are considered, the model is reduced to two independent open systems, describing a two-level atom connected to two heat reservoirs of different temperatures. In this case, the master equation is of the same form as equations (\ref{MEq},\ref{MEq.(independent)}) with $V_m=\left\vert\downarrow\rangle_m\langle\uparrow\right\vert$ and the eigenfrequency $\omega_m$. The derivation process is also given in appendix \hyperref[AppendixB]{B}.

\section{The steady-state heat current in the case of detuned coupled TLAs}
\label{sec:3}

With the master equation (\ref{MEq}), the dynamic evolution of the system can be easily obtained. However, for the thermodynamic behaviors, we are only interested in the steady state, which is determined by $\dot\rho(t)=0$ in equation (\ref{MEq}). Thus one can obtain the equations governing the steady state $\rho^S$ as
\begin{align}
\nonumber
&\mathcal{M}\vert\varrho^C\rangle=0, \\
&\varrho_{ij}^C=0,i\neq j,\label{steady Matrix}
\end{align}
where $\vert\varrho^{C}\rangle=\frac{1}{N^C}[\varrho_{11}^C,\varrho_{22}^C,\varrho_{33}^C,\varrho_{44}^C]^T$ is a column vector consisted of the diagonal elements of the steady-state density matrix $\rho^S$ with the normalization coefficient $N^C=\sum_{p=1}^4\varrho_{pp}^C$. Here, $C$ denotes CHR, $\mathcal{M}=\mathcal{M}^C=\sum_\alpha\mathcal{M}^C_\alpha$ and
\begin{equation} \mathcal{M}_\alpha^C=
\begin{pmatrix}
\mathrm{M}^{12}_\alpha+\mathrm{M}^{13}_\alpha & -\mathrm{M}^{21}_\alpha & -\mathrm{M}^{31}_\alpha & 0\\
-\mathrm{M}^{12}_\alpha & \mathrm{M}^{21}_\alpha+\mathrm{M}^{24}_\alpha & 0 & -\mathrm{M}^{42}_\alpha\\
-\mathrm{M}_\alpha^{13} & 0 & \mathrm{M}_\alpha^{31}+\mathrm{M}^{34}_\alpha & -\mathrm{M}^{43}_\alpha\\
0 & -\mathrm{M}_\alpha^{24} & -\mathrm{M}_\alpha^{34} & \mathrm{M}_\alpha^{42}+\mathrm{M}_\alpha^{43}
\end{pmatrix},\label{Mall}
\end{equation}
where $\mathrm{M}^{pq}_\alpha=-2(\breve{\mathrm{M}}^{pq}_\alpha)^2$, and
\begin{equation} \left(
\begin{smallmatrix}
\breve{\mathrm{M}}^{12}_\alpha&\breve{\mathrm{M}}^{21}_\alpha\\
\breve{\mathrm{M}}^{34}_\alpha&\breve{\mathrm{M}}^{43}_\alpha
\end{smallmatrix}\right)=\Theta^+\mathcal{J}_-
,\quad
\left(
\begin{smallmatrix}
\breve{\mathrm{M}}^{13}_\alpha&\breve{\mathrm{M}}^{31}_\alpha\\
\breve{\mathrm{M}}^{24}_\alpha&\breve{\mathrm{M}}^{42}_\alpha
\end{smallmatrix}\right)=i\sigma_y\Theta^-\sigma_x\mathcal{J}_+
\end{equation}
with 
\begin{equation}\Theta^\pm=\left(
\begin{smallmatrix}
\sin\theta_\pm&-\cos\theta_\mp\\\sin\theta_\pm&\cos\theta_\mp
\end{smallmatrix}
\right),\quad
\mathcal{J}_i=\left(
\begin{smallmatrix}
\sqrt{J^{11}_\alpha(\omega_i)}&\sqrt{J^{11}_\alpha(-\omega_i)}\\\sqrt{J^{22}_\alpha(\omega_i)}&\sqrt{J^{22}_\alpha(-\omega_i)}
\end{smallmatrix}
\right).\end{equation}
One can check that the rank of matrix $\mathcal{M}^C$ is 3, so the unique solution of equation (\ref{steady Matrix}), i.e., the steady state can be obtained as
\begin{equation}
\varrho_{ii}^{C}=\sum_{\substack{k,l=1 \\j=5-i}}^4\left\vert\epsilon_{ijkl}\right\vert(\mathrm{M}^{ki}+\mathrm{M}^{kj})\mathrm{M}^{li}\mathrm{M}^{jl}, i\in [1,4] ,\label{steady_CHR}
\end{equation}
where $\mathrm{M}^{pq}=\sum_\alpha\mathrm{M}_\alpha^{pq}$ as given in equation (\ref{Mall}) and $\epsilon_{ijkl}$ is 4-dimensional Levi-Civita symbol.

With the above steady-state density matrix, we can calculate the heat currents by the expression \cite{li2012colloquium,kosloff2013quantum}
\begin{equation}
\dot{Q}_\alpha={\rm{Tr}}[{H_S{\mathcal{L}_{\alpha}(\rho^S)}}]=\langle {\lambda}\vert {\mathcal{M}}_{\alpha}\vert \rho^S\rangle, \label{Q}
\end{equation}
where $\vert\lambda\rangle$ is a column vector consisting of the eigenvalues of $H_S$. Substituting the current system into equation (\ref{Q}), we have
\begin{align}
\nonumber
\dot{Q}_\alpha^C=&\frac{\omega_-}{N^C}[(\mathrm{M}_\alpha^{21}\varrho^{C}_{22}+\mathrm{M}_\alpha^{43}\varrho^{C}_{44})-(\mathrm{M}_\alpha^{12}\varrho^{C}_{11}+\mathrm{M}_\alpha^{34}\varrho^{C}_{33})]\\
&+\frac{\omega_+}{N^C}[(\mathrm{M}_\alpha^{31}\varrho^{C}_{33}
+\mathrm{M}_\alpha^{42}\varrho^{C}_{44})-(\mathrm{M}_\alpha^{13}\varrho^{C}_{11}+\mathrm{M}_\alpha^{24}\varrho^{C}_{22})].\label{Qtotal}
\end{align} 
\textit{If we don't consider the coupling between atoms}, the above HC in equation (\ref{Qtotal}) for $\omega_1\neq\omega_2$ will become
\begin{equation}
\dot{Q}_\alpha^{C\prime}=\sum_{m=1}^2
\omega_m\frac{\mathbbm{J}^{m-}_\alpha \mathbbm{J}^{m+}-\mathbbm{J}^{m+}_\alpha \mathbbm{J}^{m-}}{\mathbbm{J}^{m+}+\mathbbm{J}^{m-}},
\label{heatCHRg0}
\end{equation} where $\mathbbm{J}^{m\pm}=\sum_\alpha\mathbbm{J}_\alpha^{m\pm}$, $m=1,2$ with
$\mathbbm{J}_\alpha^{m\pm}=-2J_\alpha ^{mm}(\pm\omega_m)$ and the prime denotes the physical quantity in the case of decoupled TLAs. Here we have to emphasize that the case with $\omega_1=\omega_2$ will be left in the latter.

\textit{Now let's first address the IHRs}, in which case the cross dissipation $\mathcal{L}_\alpha^{mn}(m\neq n)$ won't be considered. Thus the steady-state density matrix vector as 
$\vert\varrho^{I}\rangle=\frac{1}{N^{I}}[\varrho_{11}^I,\varrho_{22}^I,\varrho_{33}^I,\varrho_{44}^I]^T$, $N^I=\sum_{p=1}^4\varrho_{pp}^I$, satisfy $\mathcal{M}^I\vert\varrho^I\rangle=0$ with the coefficient matrix
\begin{equation}
\mathcal{M}^I_\alpha=\left(
\begin{matrix}
M^{1}&-M^{3}\\
-M^{1}&M^{3}
\end{matrix}
\right)\otimes\mathbf{1}_2\\
+\mathbf{1}_2\otimes\left(
\begin{matrix}
M^{2}&-M^{4}\\
-M^{2}&M^{4}
\end{matrix}
\right),
\end{equation}
where superscript $I$ stands for IHR, $\mathbf{1}_2$ denotes 2-dimensional identity matrix and $M^p=\sum_\alpha M_\alpha^p$ with 
\begin{align}
M_\alpha^{1}=&M_\alpha^{1}(\omega_-)=-2[\sin^2\theta_+J_\alpha ^{11}(\omega_-)+\cos^2\theta_-J_\alpha ^{22}(\omega_-)],
&\quad M_\alpha^{3}=M_\alpha^{1} (-\omega_-),
\nonumber\\
M_\alpha^{2}=&M_\alpha^{2}(\omega_+)=-2[\cos^2\theta_+J_\alpha ^{11}(\omega_+)+\sin^2\theta_-J_\alpha ^{22}(\omega_+)],
&\quad M_\alpha^{4}=M_\alpha^{2} (-\omega_+).
\label{element-independent}
\end{align}
Due to the ignored crossing dissipation in the current case, all the parameters in equation (\ref{element-independent}) is related to those in equation (\ref{Mall}) as $\mathrm{M}^{12}_\alpha=\mathrm{M}^{34}_\alpha=M^1_\alpha$, $\mathrm{M}^{13}_\alpha=\mathrm{M}^{24}_\alpha=M^2_\alpha$, $\mathrm{M}^{21}_\alpha=\mathrm{M}^{43}_\alpha=M^3_\alpha$ and $\mathrm{M}^{31}_\alpha=\mathrm{M}^{42}_\alpha=M^4_\alpha$. Therefore, the steady state can be given in a simple form as
\begin{alignat}{2}
\nonumber
\varrho_{11}^I &=M^3M^4, &\quad \varrho_{22}^I &=M^1M^4,\\
\varrho_{33}^I &=M^2M^3, &\quad \varrho_{44}^I &=M^1M^2,
\label{steady_IHR}
\end{alignat}
and the heat currents given in equation (\ref{Q}) can also be simplified as 
\begin{align}
\dot{Q}_\alpha^I=&\frac{\omega_-}{N^I}[M_\alpha^3(\varrho^{I}_{22}+\varrho^{I}_{44})-M_\alpha^1(\varrho^{I}_{11}+\varrho^{I}_{33})]\nonumber\\
&+\frac{\omega_+}{N^I}[M_\alpha^4(\varrho^I_{33}+\varrho^I_{44})-M_\alpha^2(\varrho^I_{11}+\varrho^I_{22})].\label{heat_IHR}
\end{align}

\textit{If we don't consider the coupling between atoms in the case of IHRs}, the steady-state HC can be given in a simple form as $\dot{Q}_\alpha^{I\prime}=\dot{Q}_\alpha^{C\prime}$ since the two atoms with different intrinsic frequency connected to CHR don't produce crossing dissipation when $g=0$, which is consistent with the case of IHR. It is obvious that the HC for $\omega_1=\omega_2=\omega$ reads
$\dot{Q}^\prime_\alpha=\frac{2\omega}{\mathbbm{J}^{+}+\mathbbm{J}^{-}}
[\mathbbm{J}^{-}_{\alpha}\mathbbm{J}^{+}-\mathbbm{J}^{+}_{\alpha}\mathbbm{J}^{-}]
$ with $\mathbbm{J}^{\pm}=\sum_\alpha \mathbbm{J}^{\pm}_\alpha$ and $\mathbbm{J}^{m\pm}_\alpha=\mathbbm{J}^\pm_\alpha$ for $m=1,2$.

\begin{figure}
	\subfigure[ ]{\includegraphics[width=8.0 cm]{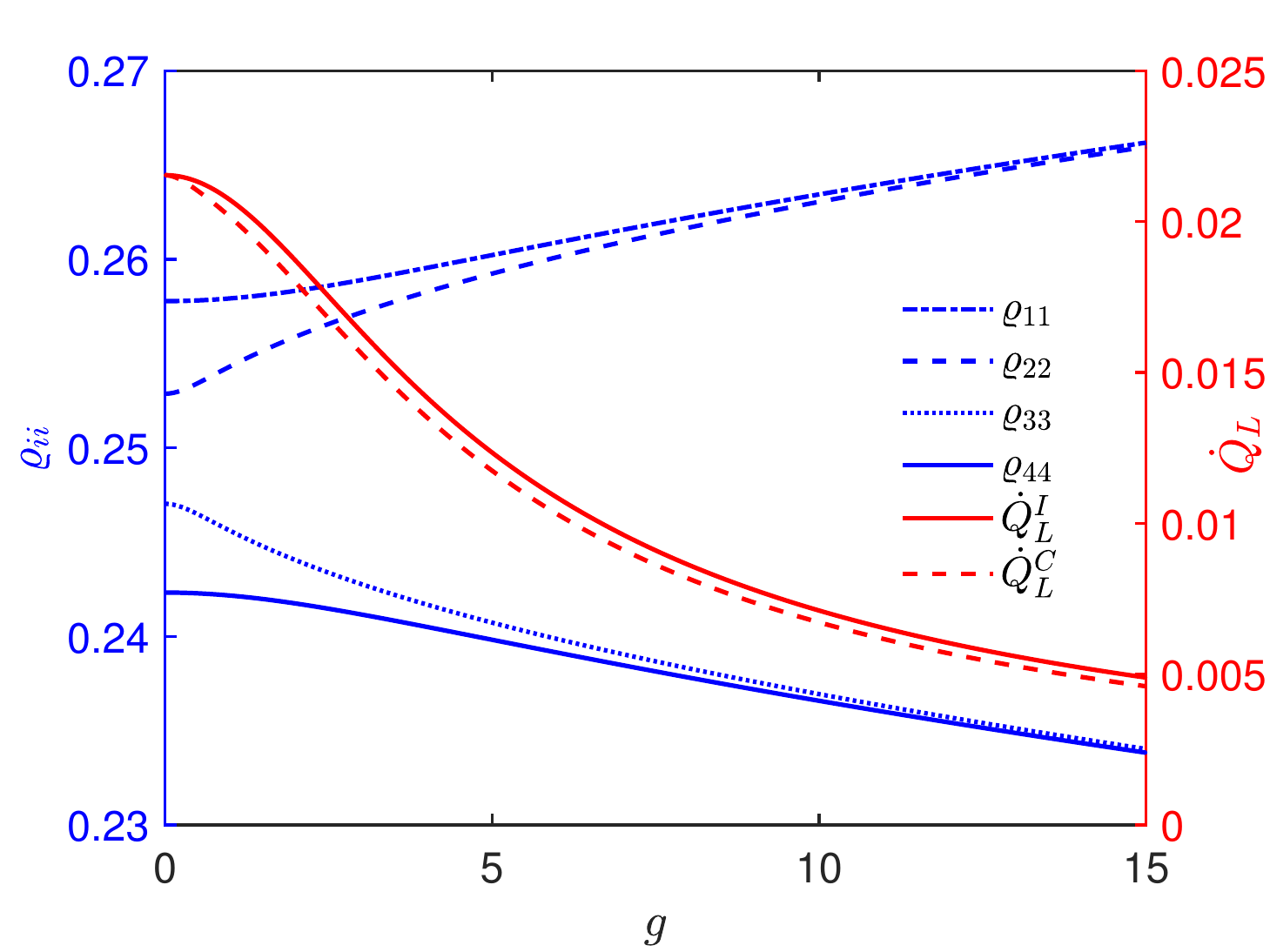}
	\label{Qandpopulation_g}}
	\subfigure[ ]{
		\includegraphics[width=8.0cm]{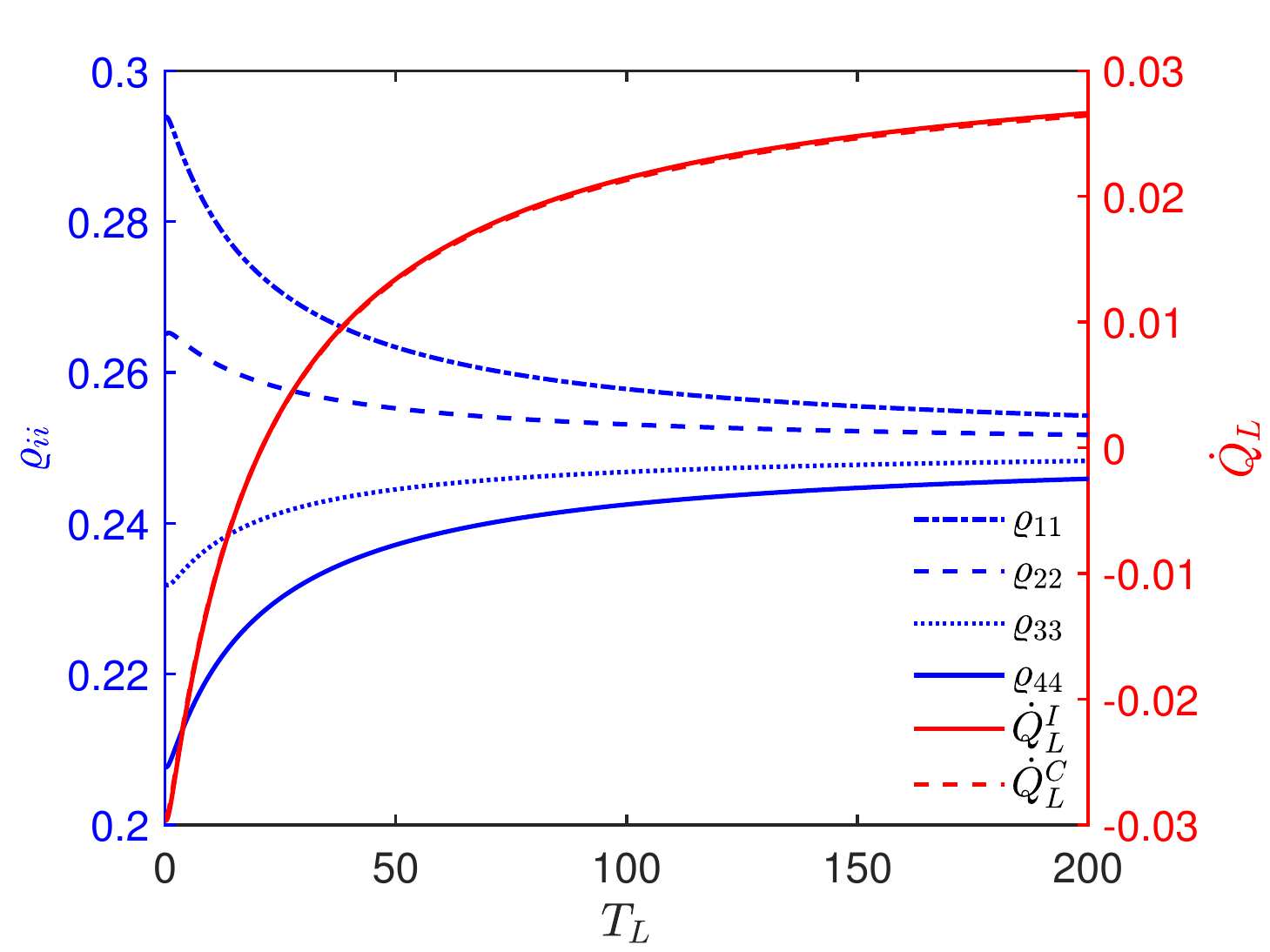}
    \label{Qandpopulation_TL}}
	\caption{HCs and populations $\varrho_{ii}$ versus atomic internal coupling strength $g$ in (a) and temperature $T_L$ in (b). The blue lines depict populations, and the red solid or dashed line represents HC of the system connecting the IHR and CHR, respectively. Here $\omega_1=3$, $\omega_2=4$, $T_R=21$ and $\gamma_-=\gamma_+=0.001\omega_1$ are fixed, and $T_L=100$ in (a) and $g=0.1\omega_1$ in (b).}
\label{HCandpop}
\end{figure}

It is necessary to mention here that when the dissipation rate meets $\gamma^{mn}_\alpha(\omega_i)=\gamma_i$, the two coupled TLSs connect to IHRs or CHRs will get the same steady state, namely, equations (\ref{steady_CHR},\ref{steady_IHR}) are identical. The specific form of the steady state is given in appendix \hyperref[AppendixC]{C}. In figure \ref{HCandpop}, we investigate the relationship between HCs $\dot{Q}_L=-\dot{Q}_R$ and steady-state populations $\varrho_{ii}$, and we find that populations of two levels with higher (or lower) energy always have the same change, increasing (decreasing) with the increase of $g$ in figure \ref{Qandpopulation_g} and decreasing (increasing) with the increase of $T_L$ in figure \ref{Qandpopulation_TL}. It also can be found that the heat currents for both CHRs and IHRs increase with the decrease (increase) of the populations of $\varrho_{11}$ and $\varrho_{22}$ ($\varrho_{33}$ and $\varrho_{44}$).

\begin{figure}
	\subfigure[ ]{\includegraphics[width=8.0 cm]{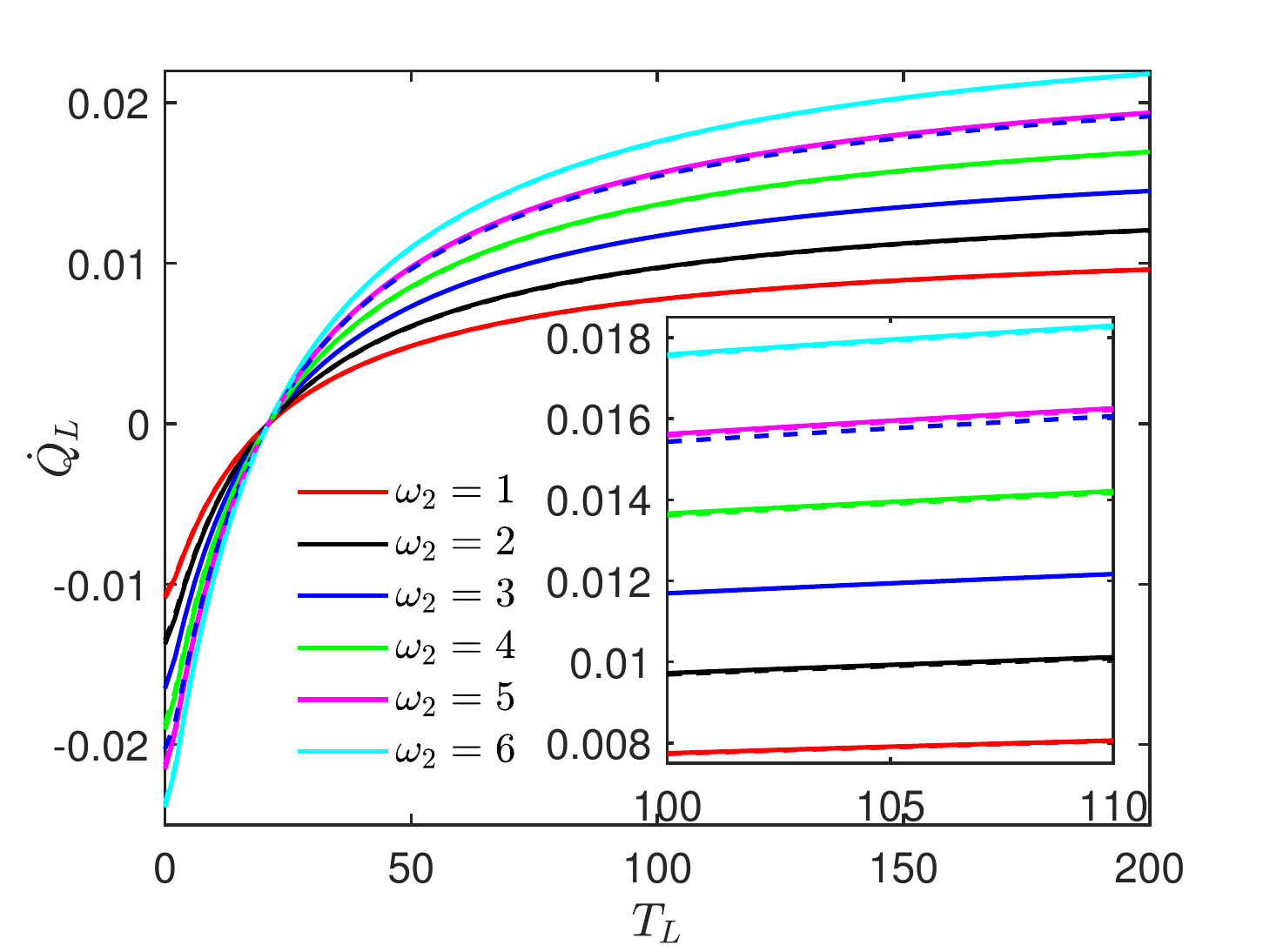}
	\label{Q_w2}}
	\subfigure[ ]{
		\includegraphics[width=8.0cm]{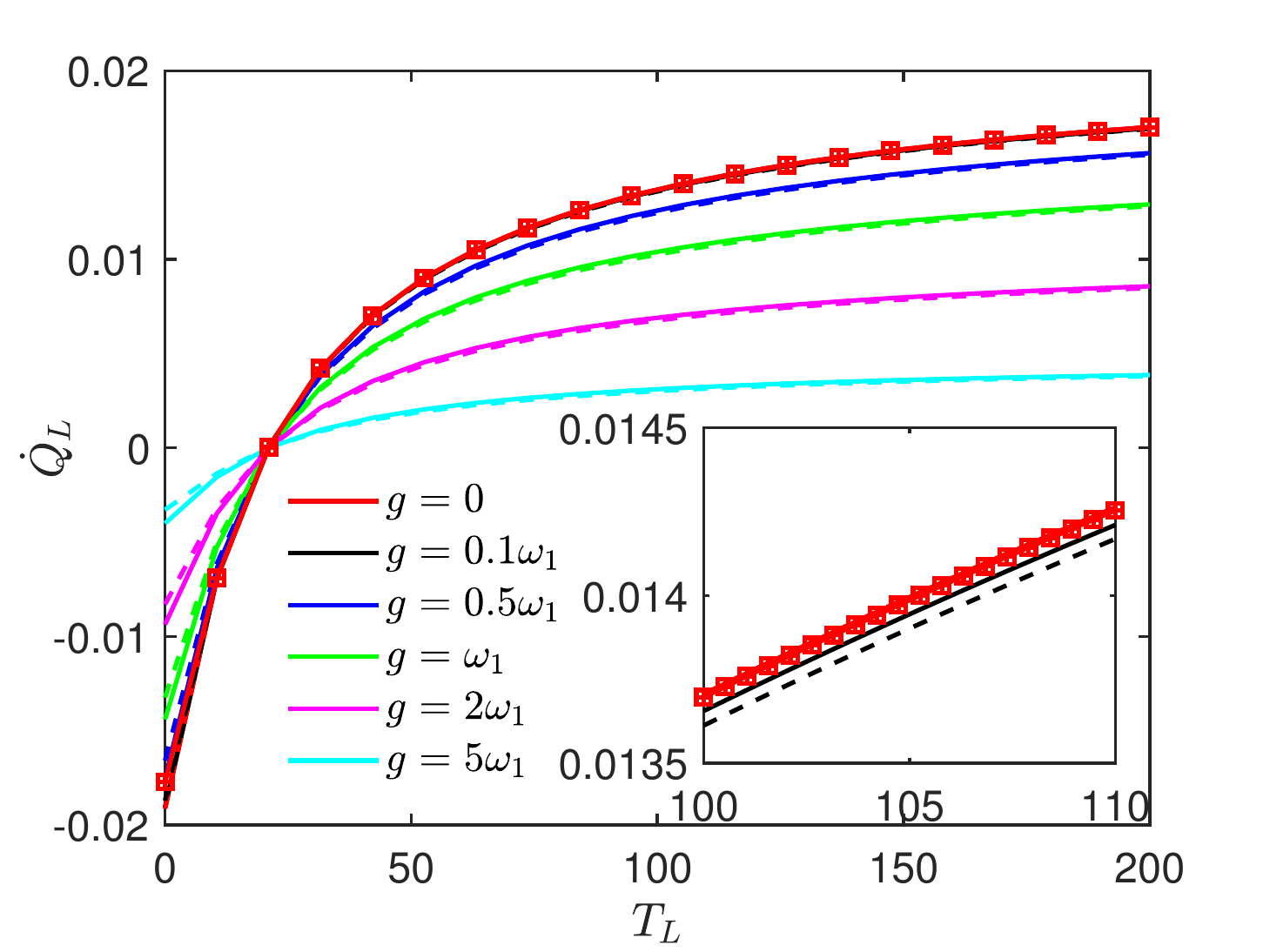}
    \label{Q_g}}\\
    
	\subfigure[ ]{\includegraphics[width=8.0 cm]{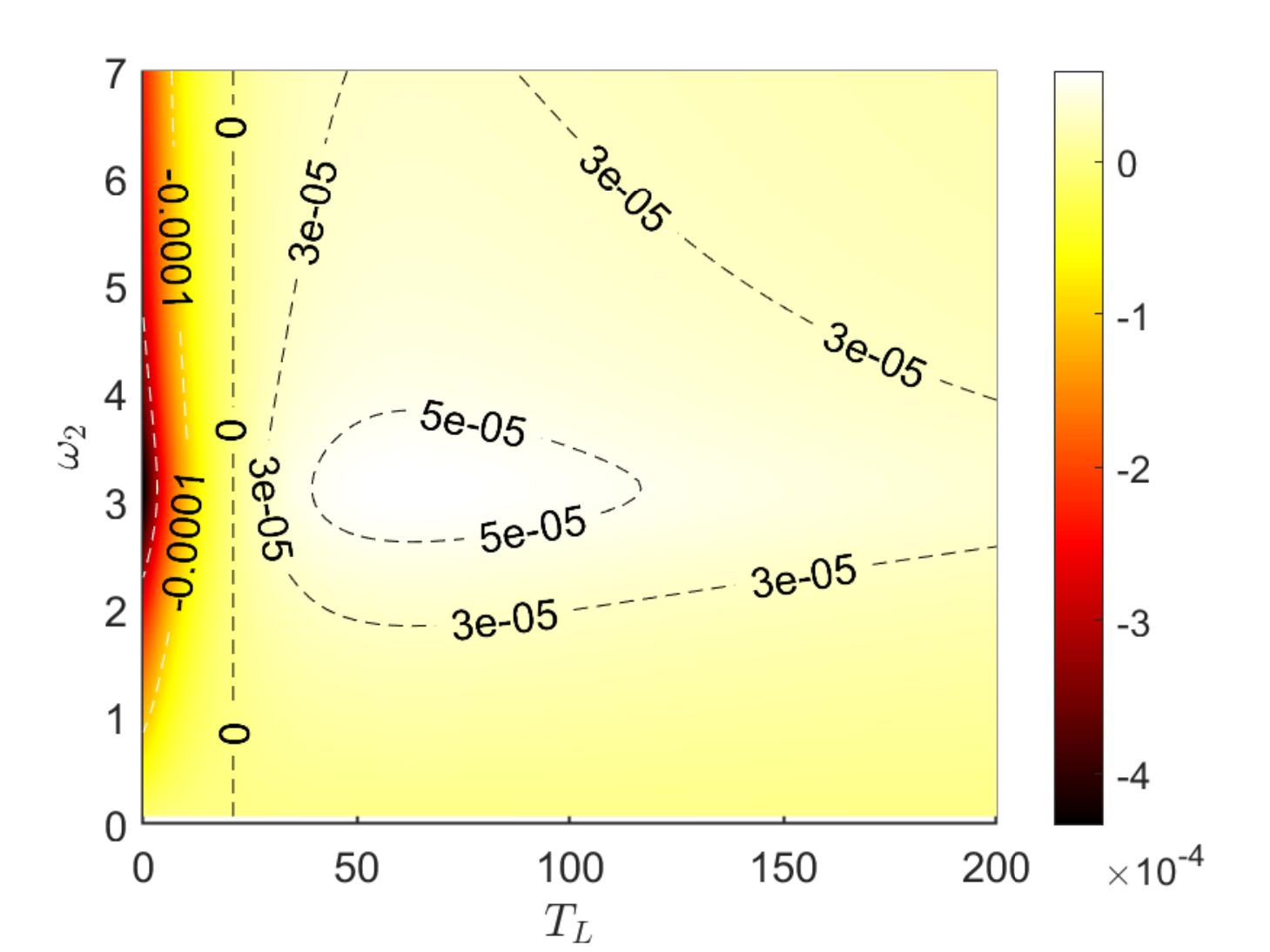}
	\label{DeltaQ_w2}}
	\subfigure[ ]{
		\includegraphics[width=8.0cm]{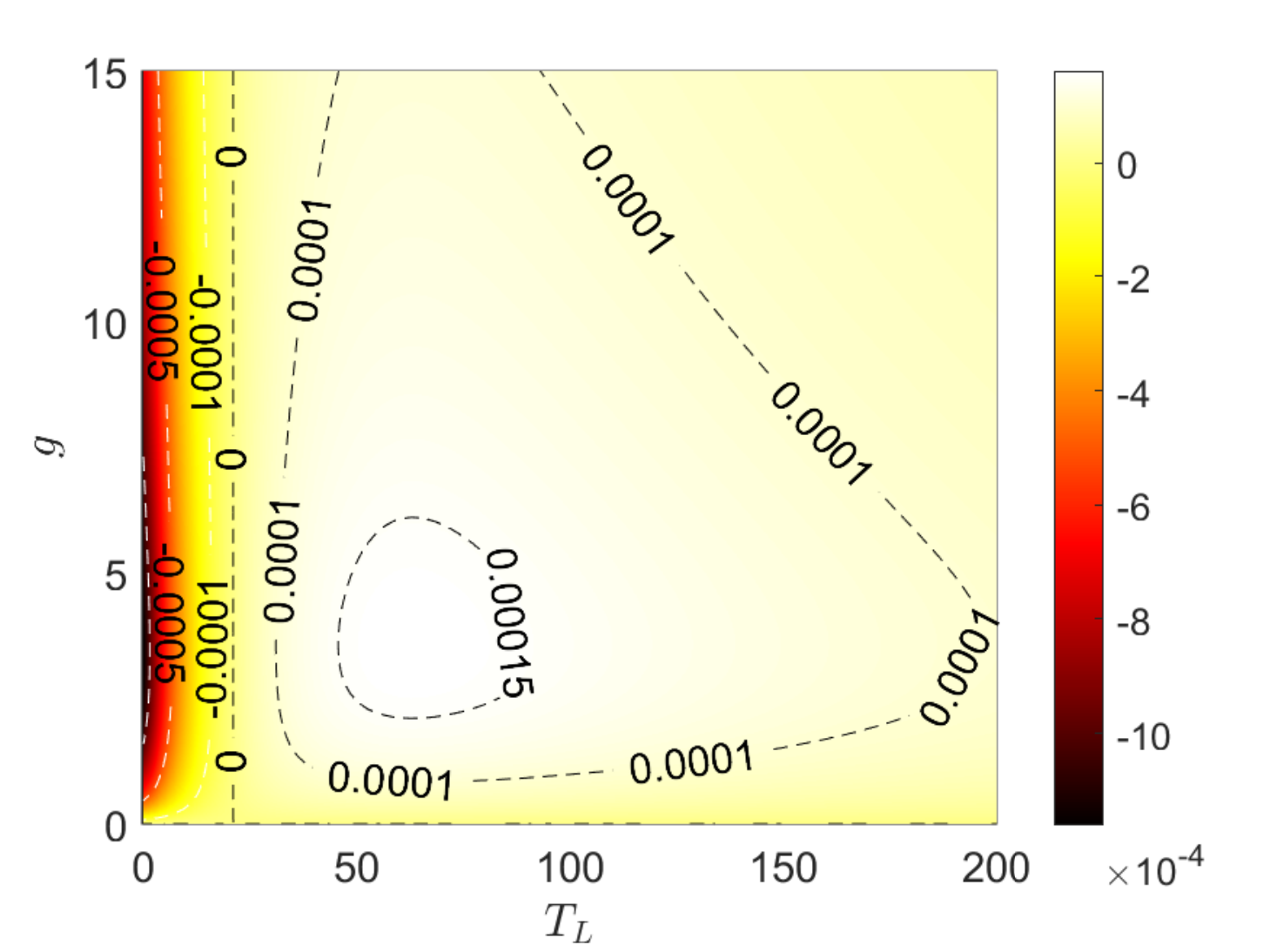}
    \label{DeltaQ_g}}
	\caption{HCs and the difference of HCs versus temperature $T_L$, atomic internal coupling strength $g$, and natural frequency $\omega_2$. The same color indicates the same parameter. (a) and (b) show the steady-state HCs $\dot{Q}_L^C$ or $\dot{Q}_L^I$ changing with temperature $T_L$, and the dashed and solid lines correspond to the heat currents $\dot{Q}_L$ with CHRs and IHRs respectively, the red plus and red square lines denote the steady-state heat current when the uncoupled TLAs are connected to IHR and CHR. (c) or (d) shows the difference $\Delta\dot{Q}_L=\dot{Q}_L^I-\dot{Q}_L^C$ changing with temperature $T_L$ and $\omega_2$ or $g$, and describing $\Delta\dot{Q}_L<0$ and $\Delta\dot{Q}_L>0$ with white and black dashed lines respectively. Here $\omega_1=3$, $\omega_2=4$, $g=0.1\omega_1$, $T_R=21$, and $\gamma_-=\gamma_+=0.001\omega_1$ are fixed.}
\label{heatcurrent_detuning_w2g}
\end{figure}

\begin{figure}
	\subfigure[ ]{\includegraphics[width=8.0 cm]{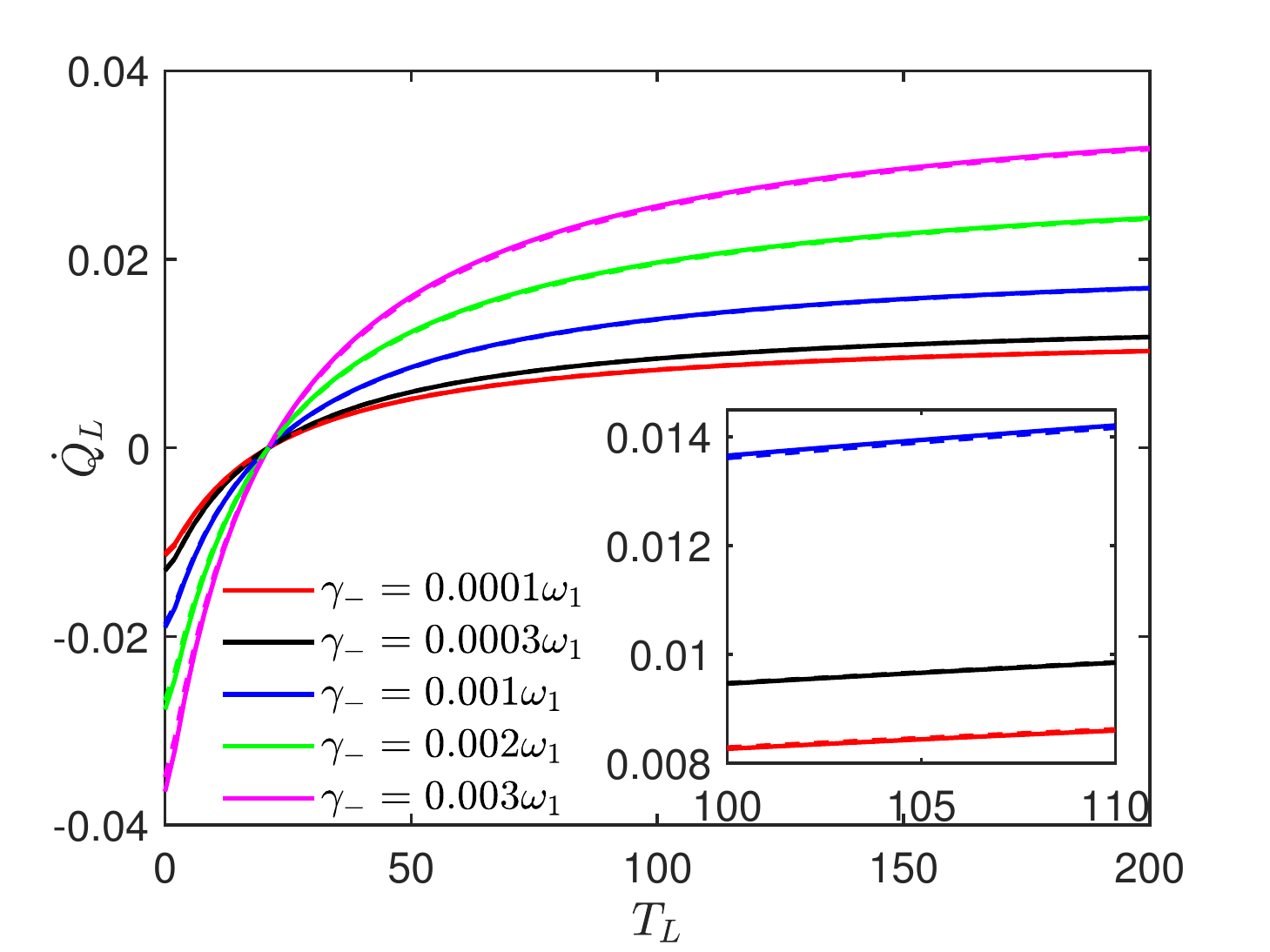}\hspace{-0.5cm}
	\label{Q_gamman}}
	\subfigure[ ]{
		\includegraphics[width=8.0cm]{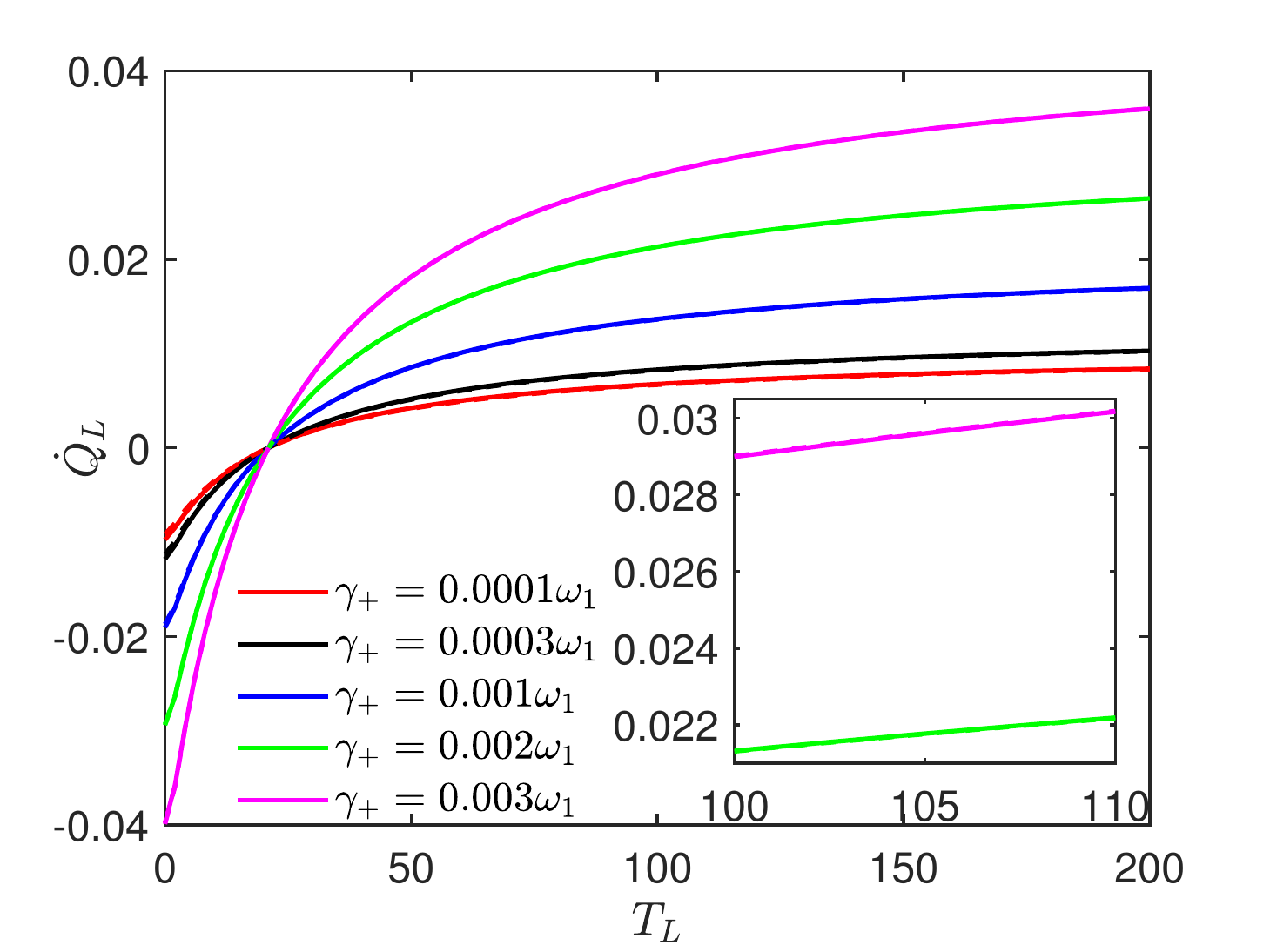}\hspace{-0.5cm}
    \label{Q_gammap}}\\
	\subfigure[ ]{
		\includegraphics[width=5cm]{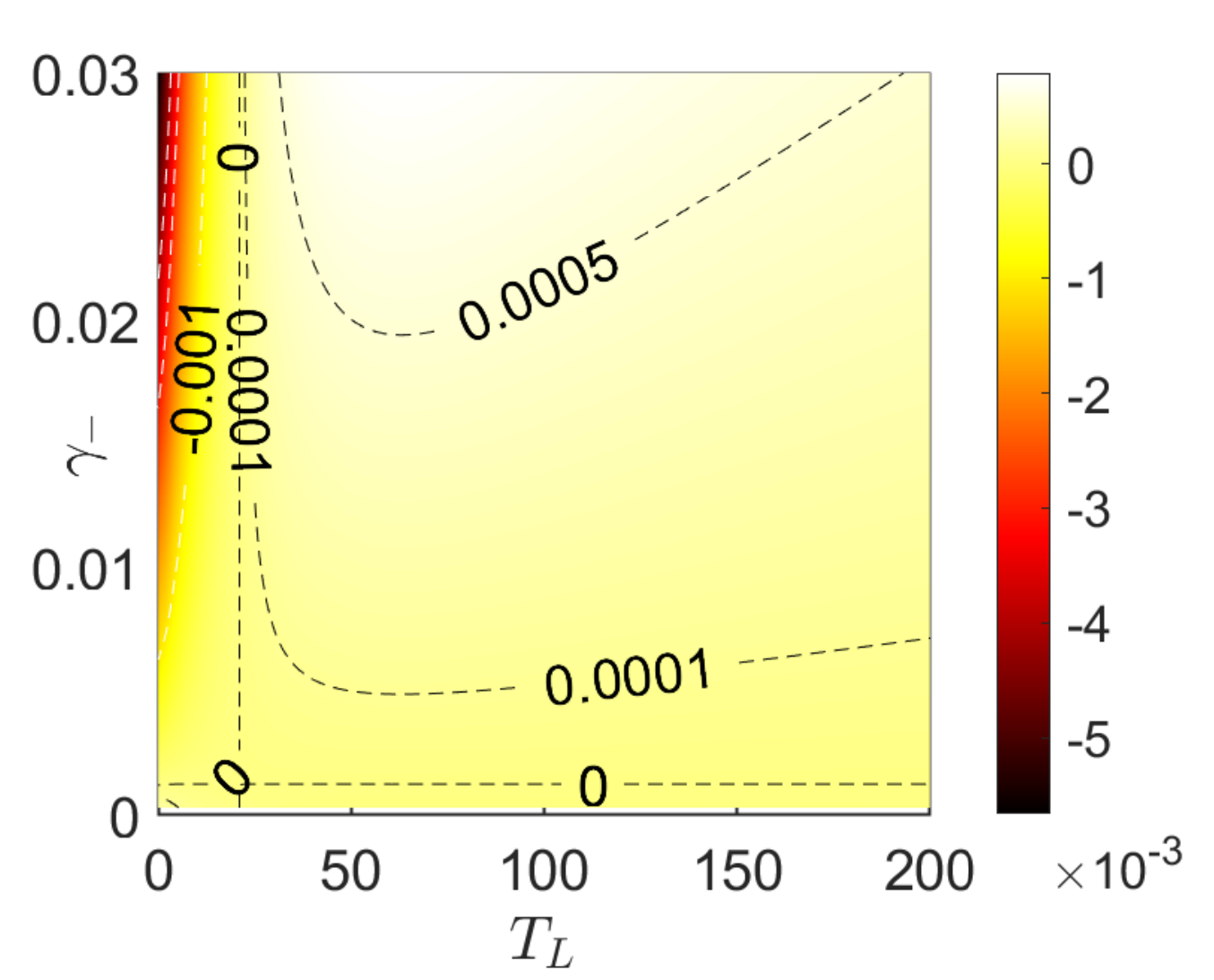}\hspace{-0.3cm}
    \label{DeltaQ_gamman}}
	\subfigure[ ]{
	\includegraphics[width=5cm]{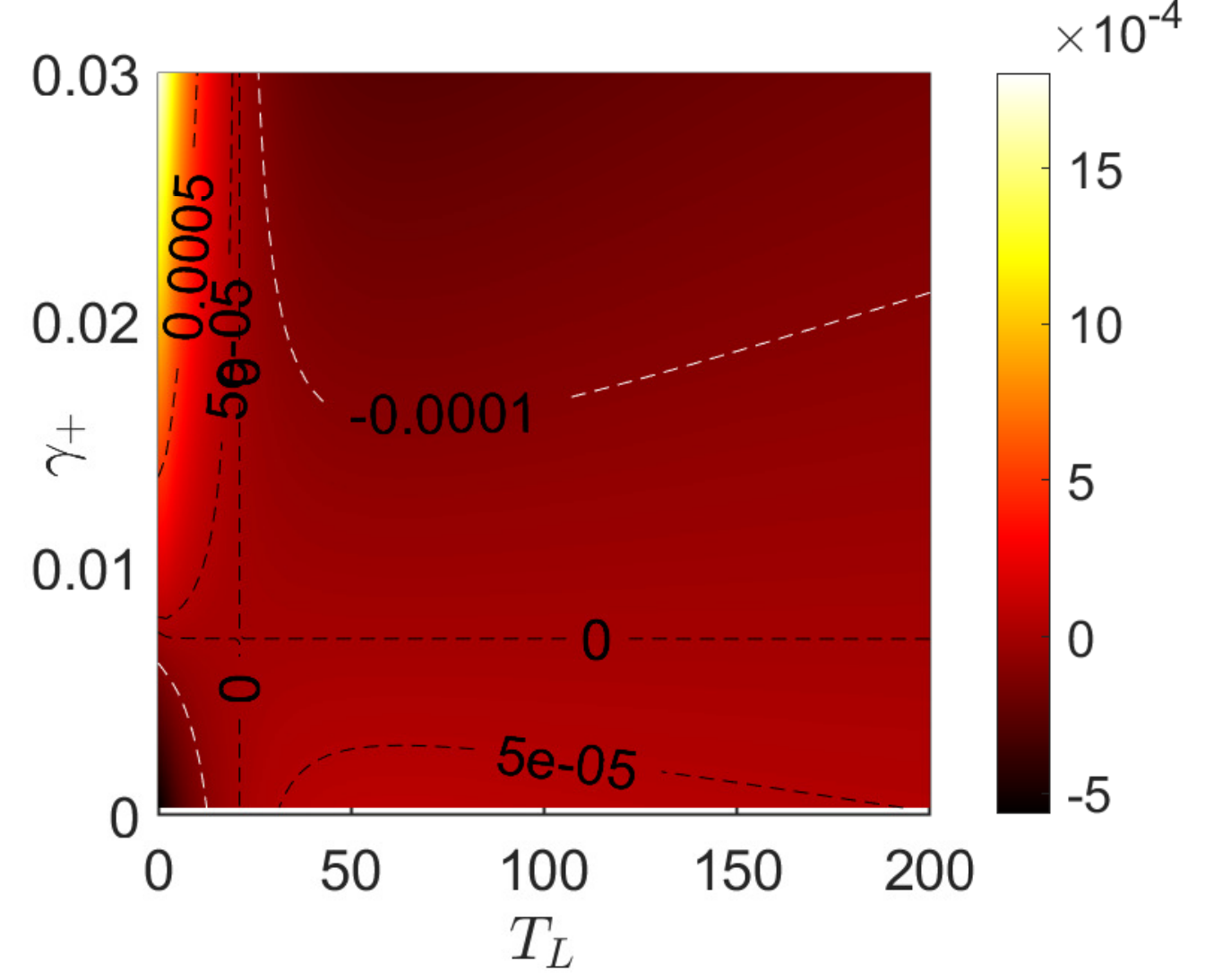}\hspace{-0.3cm}
	\label{DeltaQ_gammap}}
	\subfigure[ ]{
		\includegraphics[width=5cm]{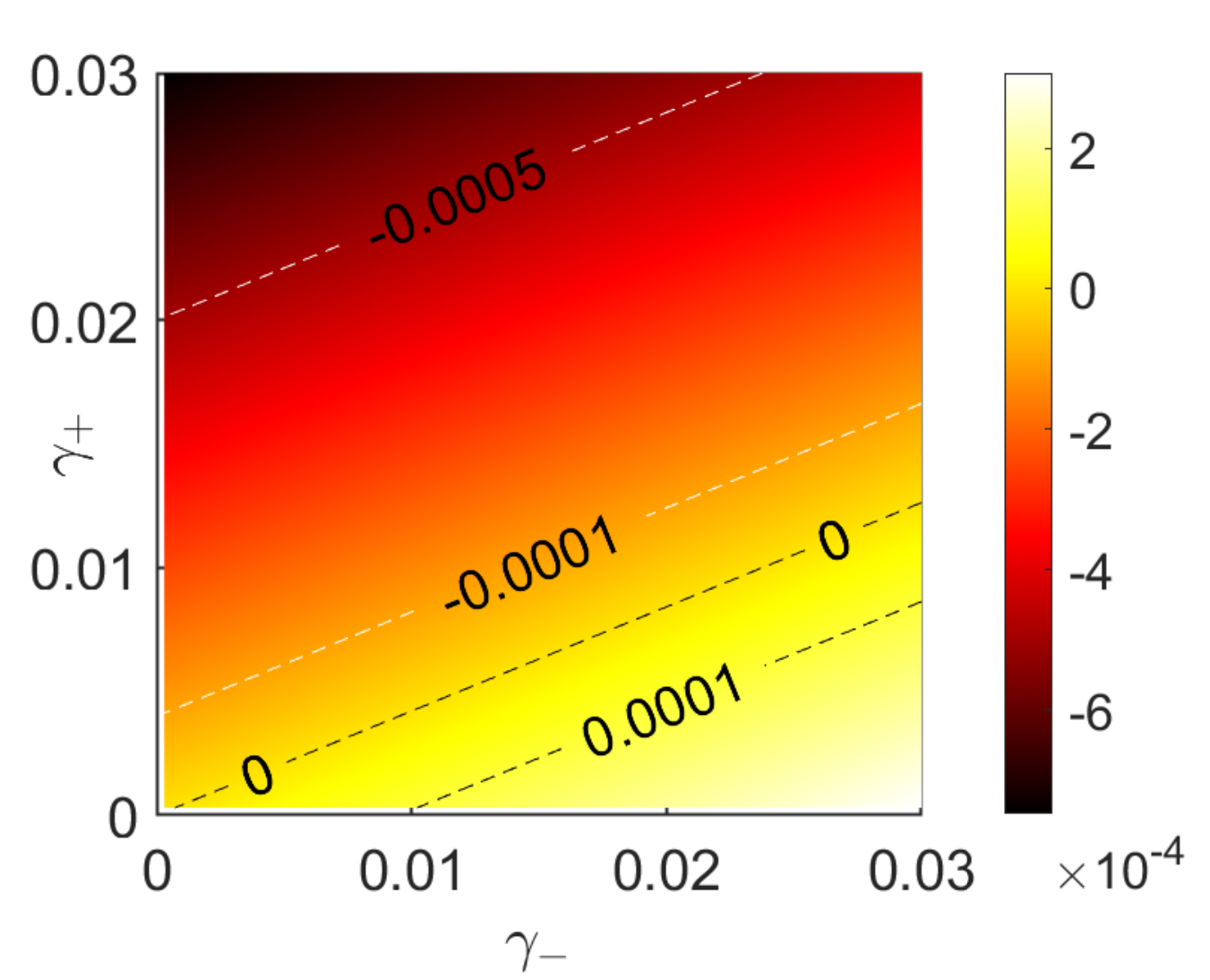}
    \label{DeltaQ_gamma}}
	\caption{HCs and the difference of HCs versus temperature $T_L$ and dissipation rate $\gamma_i$. The same color indicates the same parameter. (a) and (b) show the steady-state HCs $\dot{Q}_L^C$ or $\dot{Q}_L^I$ changing with temperature $T_L$, and the dashed and solid lines correspond to the heat currents $\dot{Q}_L$ with CHRs and IHRs, respectively, with the insets illustrating the local enlargement. (c)-(e) show the difference $\Delta\dot{Q}_L=\dot{Q}_L^I-\dot{Q}_L^C$ changing with temperature $T_L$ and $\gamma_i$, describing $\Delta\dot{Q}_L<0$ and $\Delta\dot{Q}_L>0$ with white and black dashed lines respectively. Here $\omega_1=3, \omega_2=4$, $g=0.1\omega_1$, $\gamma_i=0.001\omega_1$, $T_R=21$, and $T_L=100$ are fixed.}
\label{heatcurrent_detuning_gamma}
\end{figure}

To provide an intuitive illustration of the influence of the CHRs and the atomic coupling, we plot the HCs $\dot{Q}_L$ in figure \ref{heatcurrent_detuning_w2g} and figure \ref{heatcurrent_detuning_gamma} with various settings. Here we only consider the detuning coupling between atoms with the fixed frequency, e.g., $\omega_1\neq\omega_2$.
It is easy to find from figures \ref{Q_w2}, \ref{Q_g}, \ref{Q_gamman} and \ref{Q_gammap} that the HCs in both CHRs and IHRs become enlarged with the increasing frequency $\omega_2$ and $\gamma_i$ or with the decreasing atomic coupling $g$. 
In particular, it can be seen that the HCs with CHRs are always slightly smaller than the HCs with the IHRs in figure \ref{Q_w2} and figure \ref{Q_g}, and the more intuitive illustrations are shown in figure \ref{DeltaQ_w2} and figure \ref{DeltaQ_g}, which indicate the difference of HC for the steady state between the two types environments. Such suppression effects of CHRs, corresponding to $\Delta\dot{Q}<0$ at $T_L<T_R$ and $\Delta\dot{Q}>0$ at $T_L>T_R$, are kept, but the maximum suppression occurs when the two atoms are near resonance, i.e., $\omega_1\approx\omega_2$, or the interatomic coupling $g$ is near the super-strong coupling, i.e., $g\sim\omega_m$. Note that our numerical research indicates that the most potent suppression effect doesn't lie at the exact resonant coupling but in the region near the resonant coupling. The HCs in both CHRs and IHRs will be equal and reach the maximum if the two atoms are decoupled. In figure \ref{Q_g}, we use the red plus and red square lines to represent the steady-state HC when the uncoupled atoms connect IHR and CHR directly. Obviously, it is consistent with equation (\ref{Qtotal}) by directly putting $g=0$.

In addition, we find that in some cases the CHR can play the enhancement role as shown in figure \ref{heatcurrent_detuning_gamma}, where $\gamma_\alpha^{mn}(\omega_i)=\gamma_i$ denotes the dissipation rates corresponding to the same eigenfrequency $\omega_i$.
In figures \ref{DeltaQ_gamman}, \ref{DeltaQ_gammap}, and \ref{DeltaQ_gamma}, we fix the natural frequency of two atoms at $\omega_1=3$ and $\omega_2=4$ and mark $\Delta\dot{Q}_L<0$ and $\Delta\dot{Q}_L>0$ with white and black dashed lines, respectively.
For the case of the fixed $\gamma_+$ corresponding to the frequency $\omega_+$, the enhancement effect of CHR is only shown in the region at the lower half plane of zero contours as shown in figure \ref{DeltaQ_gamman}. However, when $\gamma_-$ is fixed in figure \ref{DeltaQ_gammap}, the part corresponding to the lower half plane of the zero contours shows the suppression effect of CHR. Figure \ref{DeltaQ_gamma} provides a clear picture of how $\Delta\dot{Q}_L$ is affected by two dissipation rates corresponding to different transitions. We found that when $\gamma_->0.42\gamma_+$, CHR always suppresses HC, on the contrary, HC is enhanced.
It is clear that $\Delta\dot{Q}_L$ is directly proportional to $\gamma_\pm$ in figure \ref{DeltaQ_gamma}. The reason is that the steady state of the system is independent not only of the type of heat reservoir but also of the dissipation rate between the system and the heat reservoir for $\gamma_\alpha^{mn}(\omega_i)=\gamma_i$, as shown in equations (\ref{rho1n},...,\ref{rho4n}). And it is proved that the effect of CHR is always suppressed under the condition $\gamma_-=\gamma_+=\gamma$.
What's more, for resonantly coupled atoms, the set of $\gamma^{mn}_\alpha(\omega_i)$ will directly change the steady-state properties of the system, as explained in the next section.

\section{The steady-state heat current in the case of resonantly coupled TLAs}
\label{sec:4}

\textit{Now we consider the resonant coupling of the two atoms}, i.e., $\omega_1=\omega_2=\omega$. One can find that the resonant or detuning coupling between atoms only affects the case of CHR, so we won't consider the case of IHR in this section. Without loss of generality, we consider the dissipation rate inconsistencies between the two atoms and the corresponding IHRs, i.e., $\gamma_{\alpha}^{11}(\omega_i)\neq \gamma_{\alpha}^{22}(\omega_i)$. In the case of resonant coupling, the diagonalized Hamiltonian can be expressed as $H_S=\sum_{l=1}^4\lambda_l\vert l\rangle_l\langle l\vert$, where
\begin{equation}
[\lambda_1,\lambda_2,\lambda_3,\lambda_4]=[-\sqrt{\omega^2+g^2},-g,g,\sqrt{\omega^2+g^2}]\label{eigval}
\end{equation}
is the eigenvalue, and
\begin{align}
\vert1\rangle&=-\sin\theta\vert\uparrow\uparrow\rangle+\cos\theta\vert\downarrow\downarrow\rangle,\\
\vert2\backslash 3\rangle&=\mp\frac{1}{\sqrt{2}}\vert\uparrow\downarrow\rangle+\frac{1}{\sqrt{2}}\vert\downarrow\uparrow\rangle,\\
\vert4\rangle&=\cos\theta\vert\uparrow\uparrow\rangle+\sin\theta\vert\downarrow\downarrow\rangle,
\end{align}
denote the corresponding eigenstates with $\sin\theta=\frac{g}{\sqrt{(\sqrt{\omega^2+g^2}+\omega)^2+g^2}}$. The eigenoperators $V_m(\omega_{m\mu})$ can be given as \begin{align}
V_{1\backslash 2}(\omega_{-})&=\sin\phi(\vert3\rangle\langle 4\vert\mp\vert1\rangle\langle 2\vert) ,\\
V_{1\backslash 2}(\omega_{+})&=\cos\phi(\vert1\rangle\langle 3\vert \pm \vert2\rangle\langle 4\vert) ,\label{eigvec}
\end{align}
where $\phi=\theta+\frac{\pi}{4}$ and $\omega_{\pm}=\sqrt{\omega^2+g^2}\pm g$.
It can be found that in the current case $\sin\phi=\sin\theta_+=\cos\theta_-$ and $\cos\phi=\sin\theta_-=\cos\theta_+$. Following a similar process as the detuning case, one can obtain the corresponding master equation and the steady state, which are anastomotic with equations (\ref{MEq},\ref{MEq.(common)},\ref{steady_CHR}). 
Similarly, one can also get the steady-state HC in the form of equation (\ref{Qtotal}) but all the relevant parameters are given by equations (\ref{eigval},$\cdots$, \ref{eigvec}).

\textit{If the coupling between the TLAs isn't considered in the case of CHR}, one will find the merging of the previous eigenoperators leads to the great change of the master equation which is explicitly derived for the same dissipation frequency $\omega$ in appendix \hyperref[AppendixB]{B}. Because of the secular approximation, two uncoupled atoms also have cross dissipation with the same heat reservoir.
In the representation $\{\left\vert1\right\rangle,\left\vert2\right\rangle,\left\vert3\right\rangle,\left\vert4\right\rangle\}$ with $\left\vert1\right\rangle=\left\vert\downarrow\downarrow\right\rangle$, $\left\vert2\right\rangle=\frac{1}{2}(\left\vert\downarrow\uparrow\right\rangle-\left\vert\uparrow\downarrow\right\rangle)$, $\left\vert3\right\rangle=\frac{1}{2}(\left\vert\uparrow\downarrow\right\rangle+\left\vert\downarrow\uparrow\right\rangle)$ and $\left\vert4\right\rangle=\left\vert\uparrow\uparrow\right\rangle$, we can solve the steady-state density matrix as the same form as equation (\ref{steady_CHR}) but the parameters therein have to be replaced by
 ${{M}^{12}={M}^{24}=\mathsf{M}^{1}}$, ${{M}^{13}={M}^{34}=\mathsf{M}^{2}}$, ${{M}^{21}={M}^{42}=\mathsf{M}^{3}}$ and ${{M}^{31}={M}^{43}=\mathsf{M}^{4}}$ with $\mathsf{M}^{p}=\sum_\alpha\mathsf{M}^{p}_\alpha$ and
\begin{align}
\nonumber
\mathsf{M}^{1 \backslash 2}_\alpha&=-[\sqrt{J_\alpha^{11}(\omega)}\mp \sqrt{J_\alpha^{22}(\omega)}]^2,\\
\mathsf{M}^{3\backslash 4}_\alpha&=-[\sqrt{J_\alpha^{11}(-\omega)}\mp \sqrt{J_\alpha^{22}(-\omega)}]^2.
\end{align}
Analogously, the HC in the current case with $g=0$ has also the same form as equation (\ref{Qtotal}) with the renewed parameters and $\omega_\pm=\omega$.

As mentioned above, no matter whether the two resonant TLAs are coupled to each other or not, we always obtain the steady state as well as the HCs with the same form, even though their values are different. This implies the similarity of the behaviors of HCs in the two cases. To verify this conclusion, we also make the numerical test as done in figures \ref{heatcurrent_detuning_w2g} and \ref{heatcurrent_detuning_gamma}. Our numerical research reveals that the HCs of the system versus various parameters indicate completely similar behaviors as figures \ref{heatcurrent_detuning_w2g} and \ref{heatcurrent_detuning_gamma} for $\gamma_{\alpha}^{11}(\omega_i)\neq \gamma_{\alpha}^{22}(\omega_i)$. That is, the CHR has a similar contribution to both the resonance and the detuning of the two TLAs. In this sense, we don't repeatedly provide the figures which are much analogous to figures \ref{heatcurrent_detuning_w2g} and \ref{heatcurrent_detuning_gamma}. However, it is important for the resonant case that the consistent dissipation rates $\gamma_{\alpha}^{11}(\omega_i)= \gamma_{\alpha}^{22}(\omega_i)$ can show quite interesting results.

\textit{Now let's turn to the consistent dissipation rates}, i.e., $\gamma_\alpha^{mn}(\omega_i)=\gamma_i$. We'd like to make further simplicity as $\gamma_\alpha^{mn}(\omega_i)=\gamma$, which doesn't change the physics. 
 In this case, the dissipators become much simpler as
\begin{align}
\nonumber
\mathcal{L}_\alpha(\rho)=&\sum_{i=\pm}J_\alpha(-\omega_i)[2V_i\rho V_i^\dagger -V_i^\dagger V_i\rho-\rho V_i^\dagger V_i]\\
&+J_\alpha(\omega_i)[2V_i^\dagger \rho V_i-V_i V_i^\dagger\rho-\rho V_i V_i^\dagger],
\end{align}
where $V_-=2\sin\phi\vert 3\rangle\langle 4\vert=\sum_{m=1}^2V_m(\omega_-)$ and $V_+=2\cos\phi\vert 1\rangle\langle 3\vert=\sum_{m=1}^2V_m(\omega_+)$. It is clear that transitions $1\leftrightarrow2$ corresponding to eigenfrequency $\omega_-$ and $2\leftrightarrow4$ corresponding to eigenfrequency $\omega_+$ do not exist, which also indirectly means that the dynamic evolution of the system will be independent of the second level under the special condition of $\gamma^{mn}_\alpha(\omega_i)=\gamma$. In addition, the matrix $\mathcal{M}^C$ in equation (\ref{steady Matrix}) governing the dynamics is also simplified as $\widetilde{\mathscr{M}}^C=\sum_\alpha \widetilde{\mathscr{M}}^C_\alpha$ with 
\[\widetilde{\mathscr{M}}^C_\alpha =\begin{pmatrix}
\widetilde{\mathsf{M}}_\alpha^1 & 0 & -\widetilde{\mathsf{M}}_\alpha^2 & 0\\
0 & 0 & 0 & 0\\
-\widetilde{\mathsf{M}}_\alpha^1 & 0 & \widetilde{\mathsf{M}}_\alpha^2+\widetilde{\mathsf{M}}_\alpha^3 & -\widetilde{\mathsf{M}}_\alpha^4\\
0 & 0 & -\widetilde{\mathsf{M}}_\alpha^3 & \widetilde{\mathsf{M}}_\alpha^4
\end{pmatrix} \]
and \begin{equation}
\widetilde{\mathsf{M}}_\alpha^{1\backslash 2}=-8\cos^2\phi J_\alpha(\pm\omega_+),\quad\widetilde{\mathsf{M}}_\alpha^{3\backslash 4}=-8\sin^2\phi J_\alpha(\pm\omega_-).\label{WXYZ}
\end{equation}
A simple calculation shows that $\widetilde{\mathscr{M}}^C$ is rank 2, which means that the steady state subject to equation (\ref{steady Matrix}) is not unique. The general solution can be expressed as
\begin{equation}
\vert\widetilde{\rho}^C\rangle=\rho_{22}\vert\widetilde{\rho}_1^C\rangle+(1-\rho_{22})\vert\widetilde{\rho}_2^C\rangle,
\label{steadystate}
\end{equation}
where the steady-state density vector $\vert\widetilde{\rho}_1^C\rangle=[0,1,0,0]^T$ and $\vert\widetilde{\rho}_2^C\rangle=\frac{1}{\widetilde{\mathsf{N}}^C}[\widetilde{\mathsf{M}}^2\widetilde{\mathsf{M}}^4,0,\widetilde{\mathsf{M}}^1\widetilde{\mathsf{M}}^4,\widetilde{\mathsf{M}}^1\widetilde{\mathsf{M}}^3]^T$, $\widetilde{\mathsf{N}}^C=\widetilde{\mathsf{M}}^2\widetilde{\mathsf{M}}^4+\widetilde{\mathsf{M}}^1\widetilde{\mathsf{M}}^4+\widetilde{\mathsf{M}}^1\widetilde{\mathsf{M}}^3$. It is clear that the steady state of the system is the mixture of two states $\vert\widetilde{\rho}_1^C\rangle$ and $\vert\widetilde{\rho}_2^C\rangle$. Thus the HC can be given as 
$\dot{\tilde{Q}}^C_\alpha=\dot{\tilde{Q}}_\alpha^{C,\max}(1-\rho_{22})$ with 
\begin{equation}
\dot{\tilde{Q}}_\alpha^{C,\max}=\frac{\omega_-\widetilde{\mathsf{M}}^1}{\widetilde{\mathsf{N}}^C}(\widetilde{\mathsf{M}}^4_\alpha \widetilde{\mathsf{M}}^3-\widetilde{\mathsf{M}}^3_\alpha \widetilde{\mathsf{M}}^4)+\frac{\omega_+\widetilde{\mathsf{M}}^4}{\widetilde{\mathsf{N}}^C}(\widetilde{\mathsf{M}}^2_\alpha \widetilde{\mathsf{M}}^1-\widetilde{\mathsf{M}}^1_\alpha \widetilde{\mathsf{M}}^2)\label{max}.
\end{equation}
The state $\vert\widetilde{\rho}_1^C\rangle$ is a 'dark state' independent of time and directly affects HCs.
In addition, one can find that the HC $\dot{\tilde{Q}}_\alpha^C$ decreases with $\rho_{22}$ increasing and $\dot{\tilde{Q}}_\alpha^C=0$ if $\rho_{22}=1$. In this case, the steady state $\vert\widetilde{\rho}^C\rangle=\vert\widetilde{\rho}_1^C\rangle$ in the original bare representation spanned by $\{\left\vert\uparrow\uparrow\right\rangle$, $\left\vert\uparrow\downarrow\right\rangle$, $\left\vert\downarrow\uparrow\right\rangle$, $\left\vert\downarrow\downarrow\right\rangle\}$ is given by $\rho=\left\vert\psi\right\rangle\left\langle\psi\right\vert$ with $\vert\psi\rangle=\frac{1}{\sqrt{2}}(\left\vert\downarrow\uparrow\right\rangle-\left\vert\uparrow\downarrow\right\rangle)$. 
When $\rho_{22}=0$, the corresponding steady state is 'residual state' $\vert\widetilde{\rho}_2^C\rangle$ and the HC will reach the maximal value as $\dot{\tilde{Q}}^{C,\max}$.

\textit{In the case of uncoupled TLAs } with $\gamma_\alpha^{mn}(\omega)=\gamma$, $\mathscr{M}^C_\alpha $ can be further simplified to $\mathscr{M}^{C\prime}_\alpha$ with
\[\widetilde{\mathscr{M}}^{C\prime}_\alpha=\begin{pmatrix}
2\mathbbm{J}^+_\alpha&0&-2\mathbbm{J}^-_\alpha&0\\
0&0&0&0\\
-2\mathbbm{J}^+_\alpha&0&2\mathbbm{J}^+_\alpha+2\mathbbm{J}^-_\alpha&-2\mathbbm{J}^-_\alpha\\
0&0&-2\mathbbm{J}^+_\alpha&2\mathbbm{J}^-_\alpha
\end{pmatrix}.\]
Similarly, one can find that the rank of $\widetilde{\mathscr{M}}^{C\prime}$ is 2, that is, the steady-state density matrix don't have unique form and the vector form can be given as $\vert{\widetilde{\rho}^{C\prime}}\rangle=\rho_{22}\vert{\widetilde{\rho}^{C\prime}}_1\rangle+(1-\rho_{22})\vert{\widetilde{\rho}^{C\prime}}_2\rangle$, where the two steady states with the vector forms as
$\vert{\widetilde{\rho}^{C\prime}}_1\rangle=[0,1,0,0]^T$ and $\vert{\widetilde{\rho}^{C\prime}}_2\rangle=\frac{1}{\widetilde{\mathsf{N}}^{C\prime}}[{\mathbbm{J}^-}^2,0,\mathbbm{J}^-\mathbbm{J}^+,{\mathbbm{J}^+}^2]$ with $\widetilde{\mathsf{N}}^{C\prime}={\mathbbm{J}^-}^2+\mathbbm{J}^-\mathbbm{J}^++{\mathbbm{J}^+}^2$.
Thus, one can also get the HC in this case as $\dot{\tilde{Q}}^{C\prime}_\alpha=\dot{\tilde{Q}}^{C\prime,max}_\alpha(1-\rho_{22})$, where
\begin{equation}
\dot{\tilde{Q}}^{C\prime,max}_\alpha=2\frac{\omega}{\widetilde{\mathsf{N}}^{C\prime}}(\mathbbm{J}^++\mathbbm{J}^-)(\mathbbm{J}^-_\alpha\mathbbm{J}^+-\mathbbm{J}^+_\alpha\mathbbm{J}^-)\label{maxg0}.
\end{equation}
Explicitly, equation (\ref{maxg0}) is consistent with equation (\ref{max}) subject to $g=0$.

Up to now, we have shown that for the resonant coupled TLAs with consistent dissipation rates as $\gamma_\alpha^{mn}(\omega_i)=\gamma_i$, the steady state of the system is not unique, which explicitly depends on the population of the second energy level $\rho_{22}$ regardless of the coupling between atoms. It is obvious that HC through the system varies from 0 to maximum with $\rho_{22}$ changing from 1 to 0. This provides us with the potential to control HCs by the $\rho_{22}$.

\section{Thermal modulator}
\label{sec:5}

From previous sections, one can find that the steady state of the system in the $H_S$ representation is directly related to the population $\rho_{22}$ since $\vert\rho_1\rangle$ does not evolve with time. Therefore, once $\rho_{22}$ is determined, the HC of the system is determined. In particular, the HC can be changed from $0$ to some maximal value with the change of $\rho_{22}$. In this sense, the HC can be effectively controlled by the population $\rho_{22}$. Thus our system can act as a thermal modulator. Next, we will give a detailed demonstration.

\begin{figure}
\begin{center}
\hspace{-2cm}
\includegraphics[width=18cm]{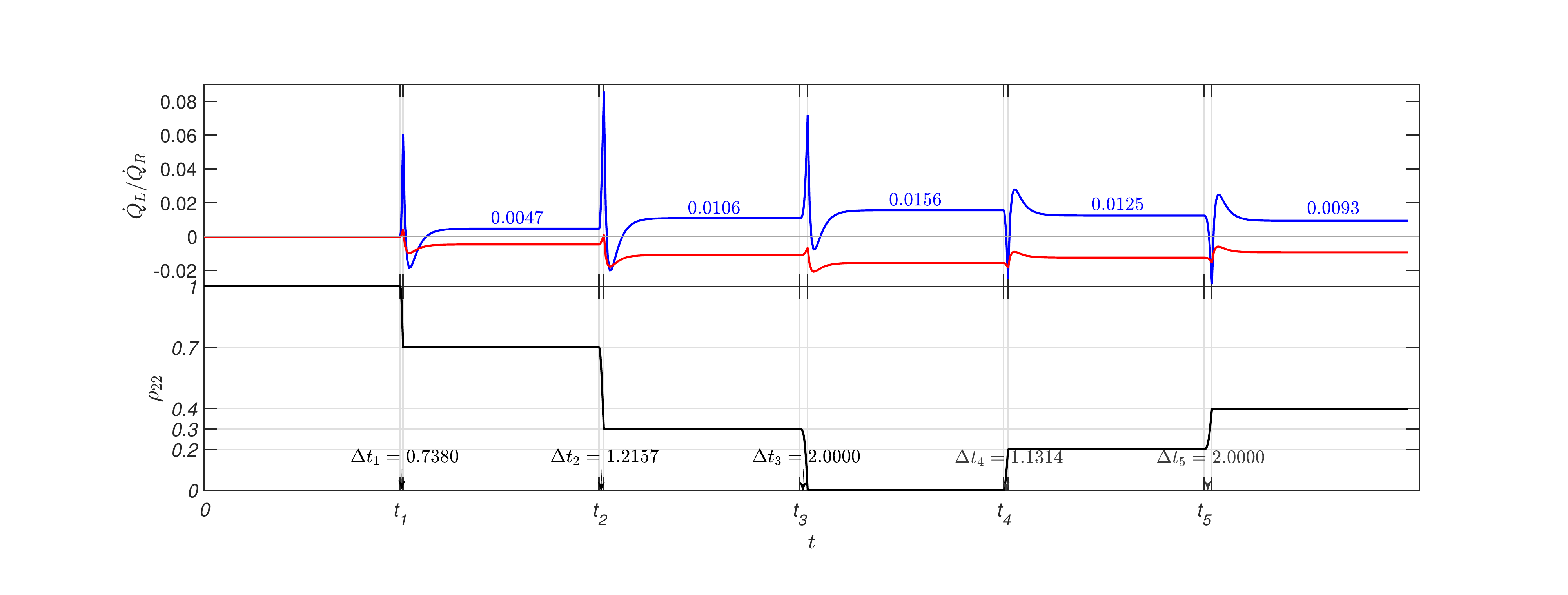}
\caption{Demonstration of the modulation of HCs $\dot{Q}_L$ and $\dot{Q}_R$. The top figure shows the transient HCs $\dot{Q}_L$ (blue) and $\dot{Q}_R$ (red) with time, and the bottom figure shows the population of the second energy level $\rho_{22}(t)$ versus the evolution time $t$. $\rho_{22}=1$ is set at the initial time. At every $t_i$, $i=1,2,\cdots,5$, a control laser field is imposed to drive the system for the time interval $\Delta t_i$. Then the control field is switched off and the system evolves freely for enough time to a steady state. In this way, the HCs can be controlled to our desired values. Here, $T_R=21$, $T_L=100$, $\omega=3$, $g=0.1\omega$, $\gamma=0.001\omega$, $\Omega_R=0.5\pi$ and the time it takes to reach steady state $t=50$.}
\label{heatcurrents_populations}
\end{center}
\end{figure}

Let's consider a laser field resonantly driving the second and another (e.g. the third) energy levels. The evolution of the system is given in appendix \hyperref[AppendixD]{D}, from which one can obtain the evolution equations of the populations as 
\begin{align}
\rho_{33}(t)&=\frac{A^-}{2}\cos(\Omega_R t)+\frac{A^+}{2},\\
\hspace{2.5cm}
\rho_{22}(t)&=-\frac{A^-}{2}\cos(\Omega_R t)+\frac{A^+}{2},
\end{align}
where $A^\pm=\rho_{33}(0)\pm \rho_{22}(0)$, $\rho_{22}(0)$ and $\rho_{33}(0)$ denote the populations of the second and the third levels of the initial state.
It can be found that $\rho_{22}(t)$ and $\rho_{33}(t)$ harmonically oscillate with time. Therefore, the population $\rho_{22}$ can be controlled by the evolution time. As an example, let's look at figure \ref{heatcurrents_populations}. The system is in the steady state at the initial moment $t_0=0$ by setting $\rho_{22}=1$ and other elements of the density matrix are zero, and it corresponds to the minimum HC $\dot{Q}_\alpha=0$, which means no heat exchanged between the two heat reservoirs. At $t_1$, we add a laser with frequency $\nu_{23}$ and Rabi frequency $\Omega_R=0.5\pi$. After the interaction time $\Delta t_1$, one gets $\rho_{22}=0.7$. Let the system evolve freely; it finally reaches the steady state with the HC $0.3$ times the maximum value. At the moment $t_2$, we let the laser drive the second and the third energy levels $\Delta t_2$ time, then $\rho_{22}=0.3$. After the free evolution for enough time, the steady-state HC becomes 0.7 times the maximum value. Then, we repeat the process; one can obtain $\rho_{22}=0$ by imposing the same laser for $\Delta t_3$ time at the moment $t_3$. The maximum HC can be obtained when the system reaches a steady state again. One can also decrease the HC by adjusting the laser driving time. But there is a little difference in the process of decreasing HC. Since $\rho_{22}$ doesn't evolve with time, the population $\rho_{22}=1$ can be reduced to $\rho_{22}=0$ by transferring the population to $\rho_{33}$. In this way, the HC can increase from zero to maximum. However, because $\rho_{11}$, $\rho_{33}$ and $\rho_{44}$ are not independent of each other in the steady state,  in order to increase the population of $\rho_{22}$ from zero to 1, we have to transfer all the other populations including $\rho_{11}$, $\rho_{33}$ and $\rho_{44}$ to $\rho_{22}$. All the transferring processes can be similarly controlled by imposing the corresponding resonantly driving laser field. As shown in figure \ref{heatcurrents_populations}, at the moment $t_4$, we adjust $\rho_{22}=0.2$ by adding the driving laser with $\Delta t_4$ interaction time. After free evolution for enough time, the HC becomes $80\%$ of the maximum value. One can also obtain $\rho_{22}=0.4$ at the moment $t_5$ by adding the laser for $\Delta t_5$ evolution time. In the steady state, HC becomes $0.6$ times the maximum value. The HC between the heat reservoirs can be controlled by repeating this process from the maximum to zero.

\section{Inverse Currents}  
\label{sec:6} 

To better understand the effects of the CHRs, in the following we will focus on the HCs subject to different dissipation subchannels. Due to the second law of thermodynamics, we know that the thermodynamic process must obey the entropy increase rate $\dot{S}=\sum_iF_iJ_i\geq 0$, where $F_i$ and $J_i$ represent the thermodynamic force and the corresponding current, respectively \cite{lieb2000fresh,kosloff2013quantum}. The generated current must be forward with the same sign as the corresponding thermodynamic force for a system with a single thermodynamic force. However, no specific rule requires that the current generated by the corresponding force must be a forward current in multiple thermodynamic force systems, as long as the resultant entropy increase rate is more significant than zero. 
A recent paper shows that inverse flow can be achieved by adjusting the temperature difference or chemical potential difference between the heat reservoirs to disturb the system's equilibrium in a one-dimensional gas model consisting of two species interacting particles in contact with two heat reservoirs \cite{wang2020inverse}.
Our system will find that the inverse heat current (IHC) can be generated if each thermodynamic force is considered a dissipative channel (DC).

According to the master equation (\ref{MEq}), we find that the dissipation of the system can be divided into direct dissipation and cross dissipation. In this sense, we can divide the DCs of each heat reservoir into the direct dissipative channels (DDC) and the crossing dissipative channels (CDC). Since HC satisfies additivity, we can separately discuss the direct heat current (DHC), the crossing heat current (CHC), and the total heat current (THC).

Let's first consider the case of section \ref{sec:3}. The matrix $\mathcal{M}^C$ can be expressed as $\mathcal{M}^C=\mathcal{M}^{d}+\mathcal{M}^{c}$, where $d$ and $c$ denote the part of direct and crossing dissipations. Based on the definition of HC given in equation (\ref{Q}), $\mathcal{M}^{d}$ and $\mathcal{M}^{c}$ can directly determine the corresponding DHC and CHC, which can be explicitly given as
\begin{align}
\dot{Q}_{\alpha}^d=&\frac{\omega_-}{N^C}[M^3_\alpha(\varrho_{22}^{C}+\varrho_{44}^{C})
-M^1_\alpha(\varrho_{11}^{C}+\varrho_{33}^{C})]\nonumber\\
&+\frac{\omega_+}{N^C}[M^4_\alpha(\varrho_{33}^{C}+\varrho_{44}^{C})
-M^2_\alpha(\varrho_{11}^{C}+\varrho_{22}^{C})],\label{DetuningQd}\\
\dot{Q}_{\alpha}^c=&\frac{\omega_-}{N^C}[\Xi^3_\alpha(\varrho_{22}^{C}-\varrho_{44}^{C})
-\Xi^1_\alpha(\varrho_{11}^{C}-\varrho_{33}^{C})]\nonumber\\
&+\frac{\omega_+}{N^C}[\Xi^4_\alpha(\varrho_{33}^{C}-\varrho_{44}^{C})-\Xi^2_\alpha(\varrho_{11}^{C}-\varrho_{22}^{C})]\label{DetuningQc},
\end{align}
where $\Xi^i=\sum_{\alpha=L,R}\Xi^i_\alpha$, $\Xi^1=M^1-\mathrm{M}^{34}=\mathrm{M}^{12}-M^1$, $\Xi^2=M^2-\mathrm{M}^{13}=\mathrm{M}^{24}-M^2$, $\Xi^3=M^3-\mathrm{M}^{43}=\mathrm{M}^{21}-M^3$ and $\Xi^4=M^4-\mathrm{M}^{31}=\mathrm{M}^{42}-M^4$.
One can easily check that $\dot{Q}_\alpha^C=\dot{Q}_\alpha^d+\dot{Q}_\alpha^c$ as given in equation (\ref{Qtotal}).
As mentioned in section \ref{sec:3}, the steady state of the system for $\gamma_\alpha^{mn}(\omega_\pm)=\gamma_\pm$ will be independent of the type of the environment. By observing equations (\ref{heat_IHR},\ref{DetuningQd}), we find that $\dot{Q}_\alpha^d=\dot{Q}_\alpha^I$ and the effect of CHR will only be reflected in the cross dissipation, i.e., $\dot{Q}_\alpha^c$.

\begin{figure}
	\subfigure[ ]{\includegraphics[width=5.5cm]{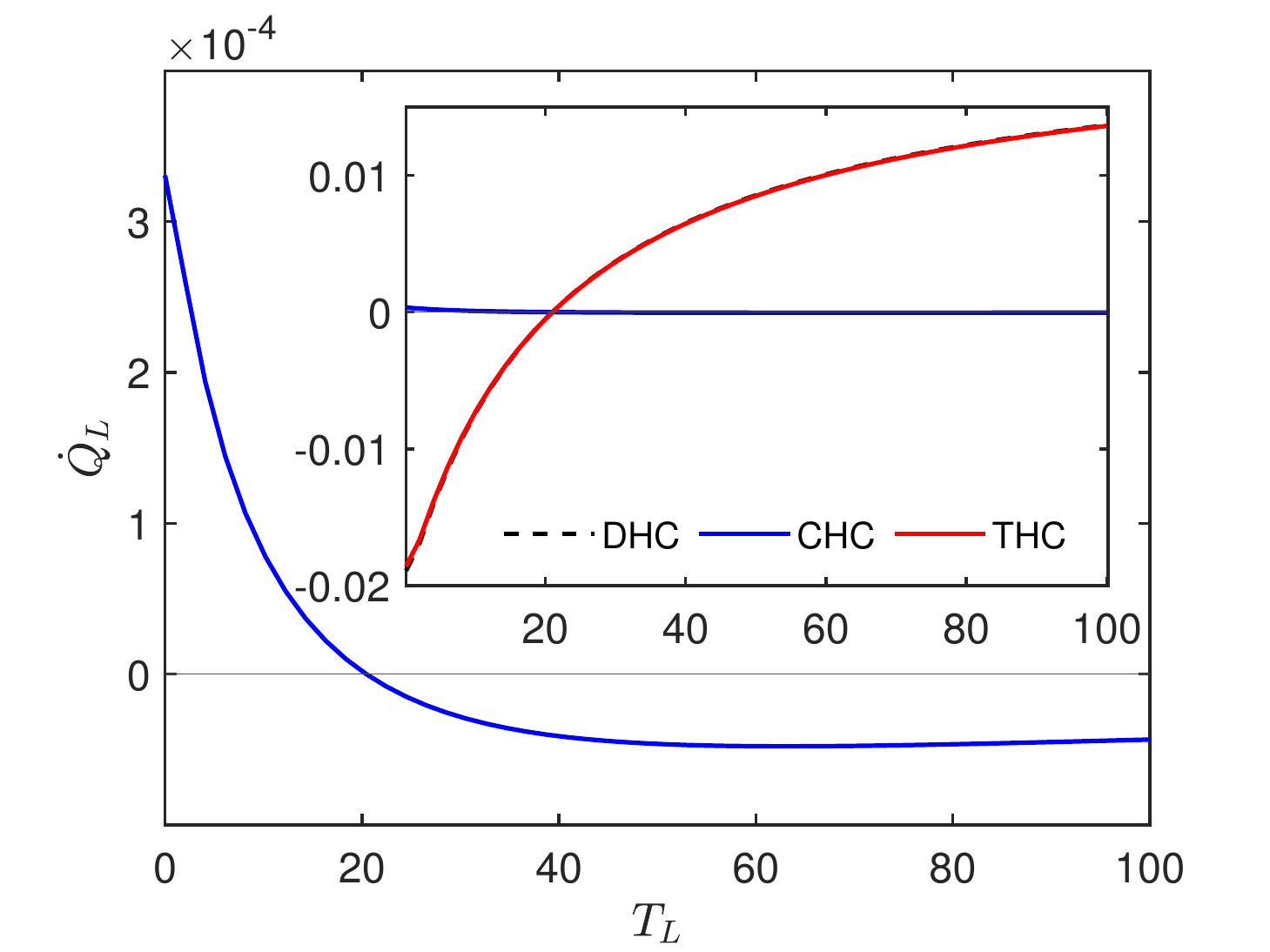}
\label{heatcurrents_of_different_channel_left}}
\hspace{-0.8cm}
	\subfigure[ ]{\includegraphics[width=5.5cm]{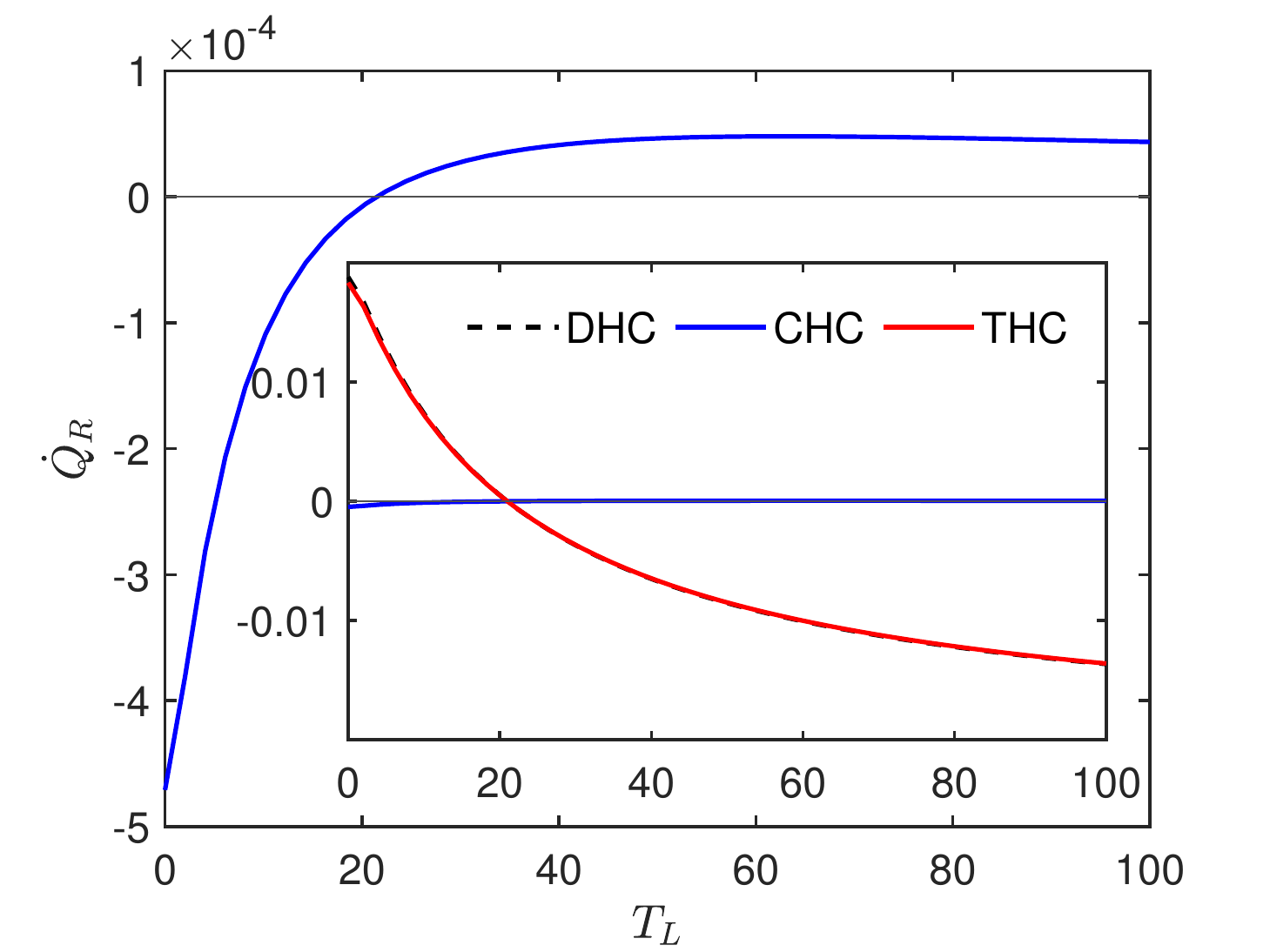}
\label{heatcurrents_of_different_channel_right}}
\hspace{-0.8cm}
   \subfigure[ ]{\includegraphics[width=5.5cm]{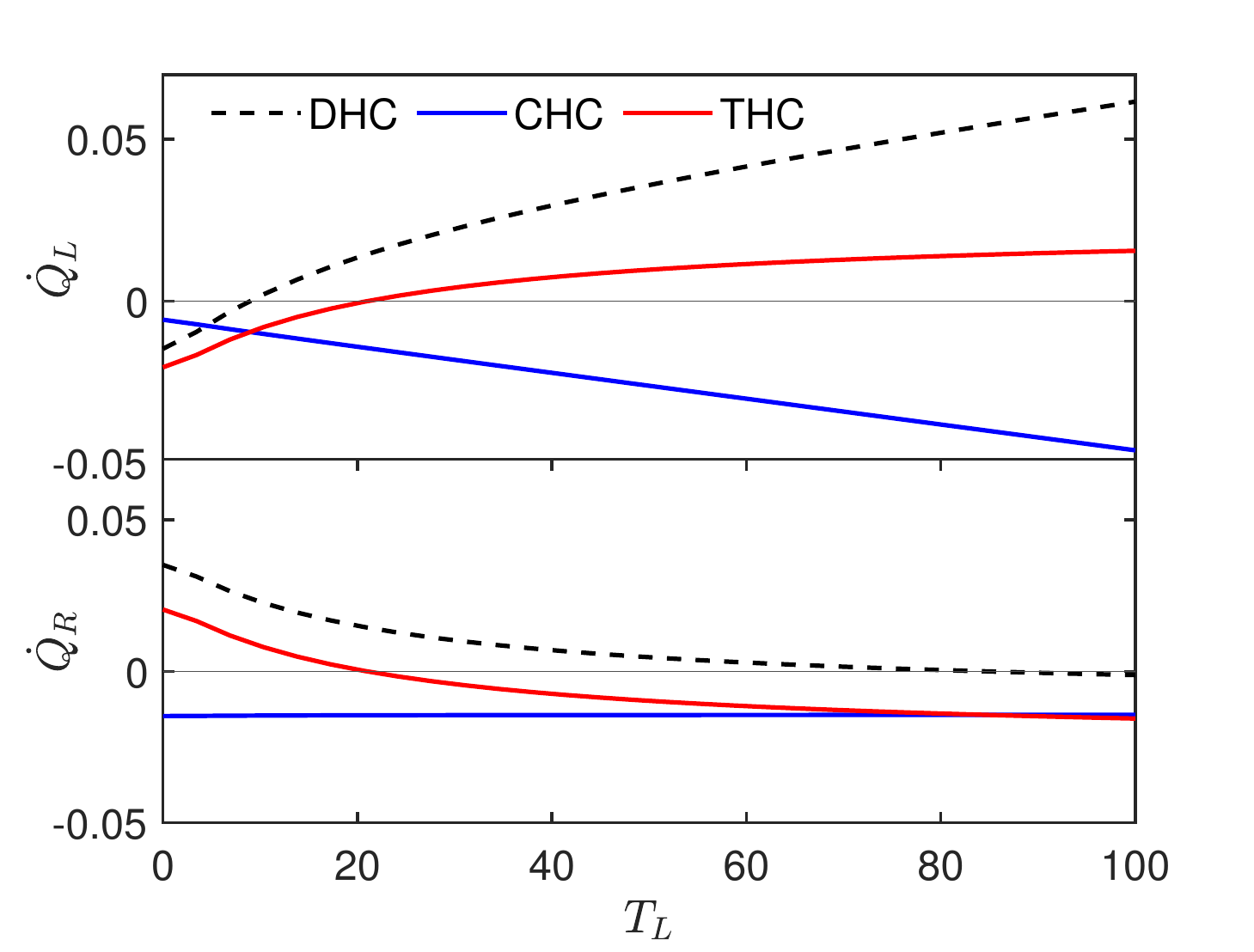}
\label{heatcurrents_of_different_channel_resonance}}
\caption{Various HCs versus $T_L$. The red solid, black dashed, and blue solid lines represent the THC, the DHC, and the CHC, respectively. The CHCs with the enlarged view in the background in (a) and (b) are opposite to the direction of the THC and hence indicate the presence of IHC. Here, $\omega_1=3$, $\omega_2=4$, $g=0.1\omega_1$, $\gamma_-=\gamma_+=0.001\omega$ and $T_R=21$. The top and the bottom figures in (c) correspond to the heat currents related to the left and the right reservoirs, respectively. The non-consistency with the direction (or sign) of the THC indicates the presence of the IHC in the channel. Here, $T_R=21$, $\omega=3$, $g=0.1\omega$, $\gamma=0.001\omega$.}
\label{heatcurrents_of_different_channel}
\end{figure}

To give an intuitive illustration of the HCs, we plot the HCs with $T_L$ in figure \ref{heatcurrents_of_different_channel}.
Figure \ref{heatcurrents_of_different_channel_left} and figure \ref{heatcurrents_of_different_channel_right} show the left and the right HCs in the steady state, where the black dashed, blue solid, and solid red lines, respectively, represent the DHC, CHC, and THC. The backgrounds in these two figures describe the CHC versus $T_L$. Interestingly, the CHC is opposite of the THC. Namely, the inverse heat current is present and inhibits the THC, which also gives a direct interpretation of why $\dot{Q}_\alpha^C$ is slightly less than $\dot{Q}_\alpha^I$ in this case.

Similarly, in the case of section \ref{sec:4}, we can also divide the matrix $\widetilde{\mathscr{M}}^C$ into direct and cross parts and get the expressions of DHCs and CHCs are
\begin{align}
\nonumber
\dot{\tilde{Q}}_\alpha^d&=\frac{1-\rho_{22}}{2\widetilde{\mathsf{N}}^C}\{\omega_-[\widetilde{\mathsf{M}}^4_\alpha \widetilde{\mathsf{M}}^1\widetilde{\mathsf{M}}^3-\widetilde{\mathsf{M}}^3_\alpha \widetilde{\mathsf{M}}^4(\widetilde{\mathsf{M}}^1+\widetilde{\mathsf{M}}^2)]-\omega_+[\widetilde{\mathsf{M}}^1_\alpha \widetilde{\mathsf{M}}^2\widetilde{\mathsf{M}}^4-\widetilde{\mathsf{M}}^2_\alpha \widetilde{\mathsf{M}}^1(\widetilde{\mathsf{M}}^4+\widetilde{\mathsf{M}}^3)]\}\\
&+\frac{\rho_{22}}{2}(\omega_-\widetilde{\mathsf{M}}^4_\alpha-\omega_+\widetilde{\mathsf{M}}^1_\alpha),\label{ResonantQd}\\
\nonumber
\dot{\tilde{Q}}_\alpha^c&=\frac{1-\rho_{22}}{2\widetilde{\mathsf{N}}^C}\{\omega_-[\widetilde{\mathsf{M}}^4_\alpha \widetilde{\mathsf{M}}^1\widetilde{\mathsf{M}}^3-\widetilde{\mathsf{M}}^3_\alpha \widetilde{\mathsf{M}}^4(\widetilde{\mathsf{M}}^1-\widetilde{\mathsf{M}}^2)]-\omega_+[\widetilde{\mathsf{M}}^1_\alpha \widetilde{\mathsf{M}}^2\widetilde{\mathsf{M}}^4-\widetilde{\mathsf{M}}^2_\alpha \widetilde{\mathsf{M}}^1(\widetilde{\mathsf{M}}^4-\widetilde{\mathsf{M}}^3)]\}\\
&-\frac{\rho_{22}}{2}(\omega_-\widetilde{\mathsf{M}}^4_\alpha-\omega_+\widetilde{\mathsf{M}}^1_\alpha)\label{ResonantQc},
\end{align}
where $\omega_\pm=\sqrt{\omega^2+g^2}\pm g$. As shown in equations (\ref{ResonantQd},\ref{ResonantQc}), there is $\dot{\tilde{Q}}_\alpha^d=-\dot{\tilde{Q}}_\alpha^c$ when $\rho_{22}=1$, that is, CHC and IHC are equal in magnitude and opposite in direction, so that there is no heat exchange between two reservoirs.

The intuitive illustration is shown in figure \ref{heatcurrents_of_different_channel_resonance}. Since the steady-state HC is not unique, we only consider the maximum HC here.
The upper and lower parts in figure \ref{heatcurrents_of_different_channel_resonance} correspond to the HC related to the left and right reservoirs, respectively. The blue, black, and red solid lines represent the DHC, the CHC, and the THC. The THC describes that the heat flows from the high-temperature to the low-temperature reservoirs, so the DHC or CHC with the opposite sign to the THC means that the IHC is generated. It is evident in figure \ref{heatcurrents_of_different_channel_resonance} that the IHCs are present related to both the left and right heat reservoirs.

It is counter-intuitive that HC spontaneously flows from the hot to the cold. This is a beautiful phenomenon due to the fantastic features of quantum mechanics. It can be understood as follows. The IHC is only a THC sub-HC, which perfectly obeys intuition. The macroscopic HC is further the statistical average of the THCs, so it is impossible to realize the spontaneous IHC macroscopically. 

\section{Concurrence of assistance }
\label{sec:7}

To further explore the quantum features of the system, we'd like to study the concurrence of assistance (COA)\cite{2003Local}, which is defined for a two-qubit density matrix $\rho$ as
\begin{equation} 
C_a(\rho)=\sum_{k=1}^4\sigma_k,\label{coa}
\end{equation} where $\sigma_k$ are the eigenvalues of the Hermitian matrix $R\equiv\sqrt{\sqrt{\rho}\tilde{\rho}\sqrt{\rho}}$ with $\tilde{\rho}=(\sigma_y\bigotimes\sigma_y)\rho^\ast(\sigma_y\bigotimes\sigma_y)$ and $\sigma_y$ denoting the Pauli matrix.

Considering the density matrix given in equation (\ref{steady_CHR}) in the bare-basis representation, one can obtain the steady-state density matrix as \[ \rho^{D}=\begin{pmatrix}
 D_D&0&0&K_D\\
 0&E_D&L_D&0\\
 0&L_D&F_D&0\\
 K_D&0&0&J_D
 \end{pmatrix}, \]
and the specific expressions of these elements $\Lambda_D=\frac{1}{N^C}\Lambda_D^\prime$, $\Lambda=\{D,E,F,J,K,L$\}, are
\begin{align}
\nonumber
D_D^\prime&=\sin^2\theta_s\varrho_{11}^C+\cos^2\theta_s\varrho_{44}^C,\\
\nonumber
E_D^\prime&=\sin^2\theta_d\varrho_{22}^C+\cos^2\theta_d\varrho_{33}^C,\\
\nonumber
F_D^\prime&=\cos^2\theta_d\varrho_{22}^C+\sin^2\theta_d\varrho_{33}^C,\\
\nonumber
J_D^\prime&=\cos^2\theta_s\varrho_{11}^C+\sin^2\theta_s\varrho_{44}^C,\\
\nonumber
K_D^\prime&=\sin\theta_s\cos\theta_s(-\varrho_{11}^C+\varrho_{44}^C),\\
L_D^\prime&=\sin\theta_d\cos\theta_d(-\varrho_{22}^C+\varrho_{33}^C).
\end{align}
The subscript or superscript $D$ denotes the detuning case.
Thus the steady-state COA can be given by
\begin{align}
C_a^{D}&=2\sqrt{D_DJ_D}+2\sqrt{E_DF_D}\nonumber\\
&=\frac{2}{N^C}(\sqrt{h_s(\varrho_{11}^C-\varrho_{44}^C)^2+\varrho_{11}^C\varrho_{44}^C}+\sqrt{h_d(\varrho_{22}^C-\varrho_{33}^C)^2+\varrho_{22}^C\varrho_{33}^C}),
\end{align}
with $h_\nu=\sin^2\theta_\nu\cos^2\theta_\nu$, $\nu=s,d$.
\begin{figure}
	\subfigure[ ]{
		\includegraphics[width=8.0cm]{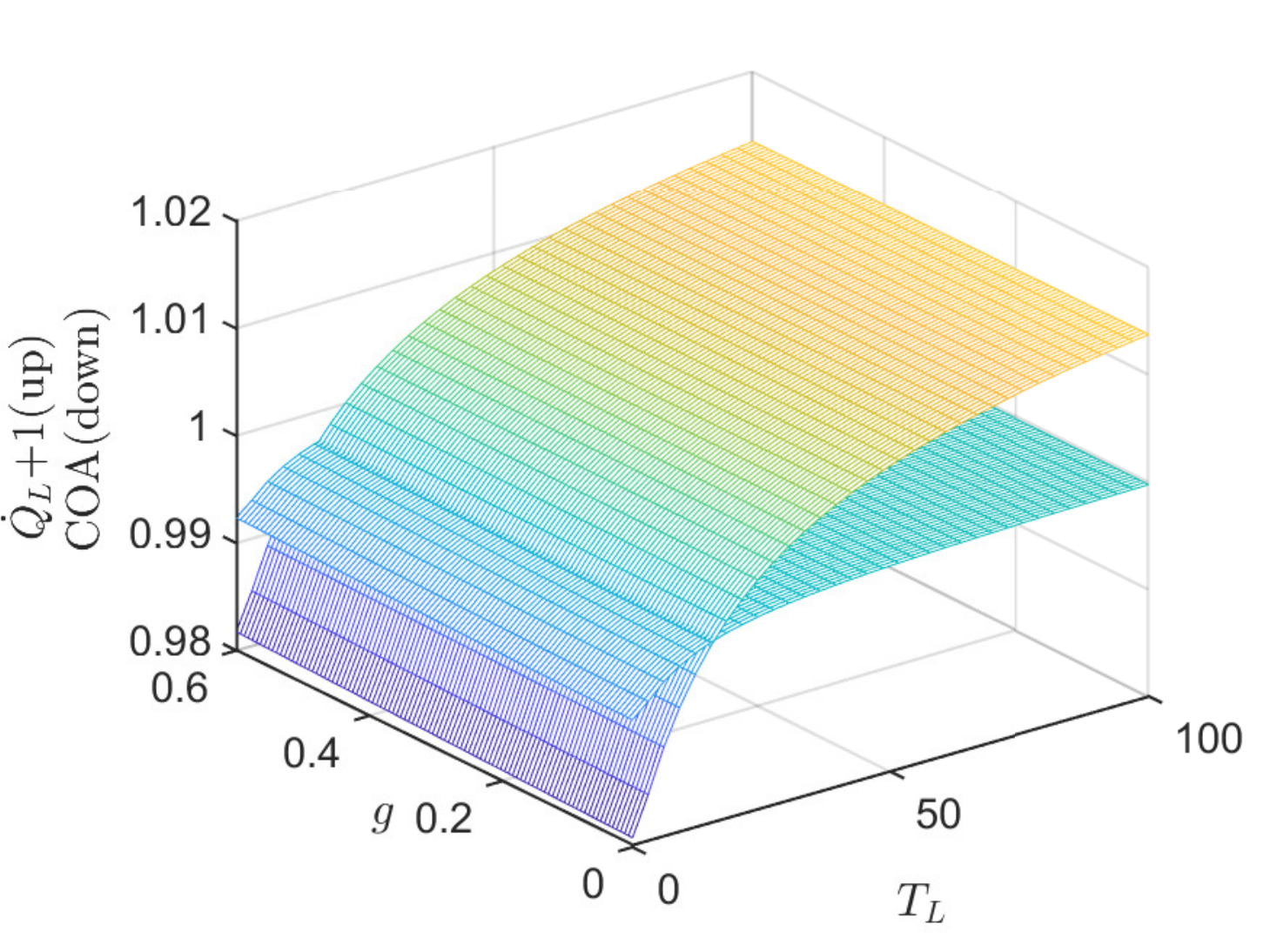}
\label{COA_heatcurrent_withTL_detuning}}
	\subfigure[ ]{\includegraphics[width=8.0 cm]{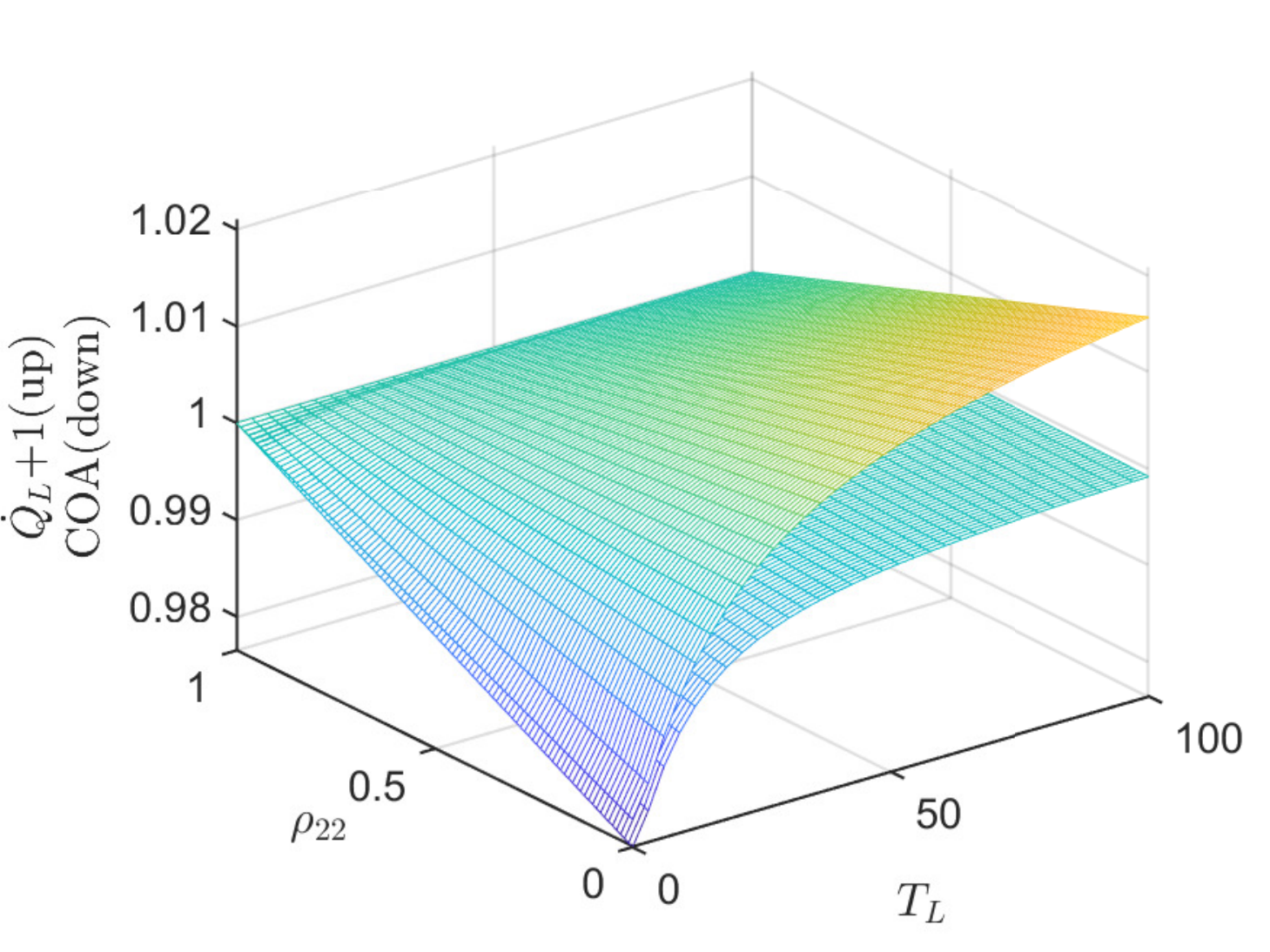}
	\label{withTL}}
	\caption{Steady-state COA and HC in both detuning and resonant cases with $T_R=21$. In (a), $\omega_1=3$, $\omega_2=4$, $T_R=21$ and $\gamma=0.001\omega_1$. In (b), $\omega_1=\omega_2=\omega=3$, $g=0.1\omega$, $T_R=21$ and $\gamma=0.001\omega$.}
\label{COA_heatcurrent_detuning}
\end{figure}

In the resonant case, we will have to consider the density matrix (\ref{steadystate}). Similarly, one can rewrite the density matrix in the bare-basis representation as
 \[ \rho^{R}=\begin{pmatrix}
 D_R&0&0&K_R\\
 0&E_R&L_R&0\\
 0&L_R&E_R&0\\
 K_R&0&0&J_R
 \end{pmatrix}, \]
where
\begin{align}
D_R&=\frac{1-\rho_{22}}{\widetilde{\mathsf{N}}^C}({\widetilde{\mathsf{M}}}^2{\widetilde{\mathsf{M}}}^4\sin^2\theta+{\widetilde{\mathsf{M}}}^1{\widetilde{\mathsf{M}}}^3\cos^2\theta),
\nonumber\\
E_R&=\frac{1-\rho_{22}}{2\widetilde{\mathsf{N}}^C}{\widetilde{\mathsf{M}}}^1{\widetilde{\mathsf{M}}}^4+\frac{\rho_{22}}{2},\nonumber\\
J_R&=\frac{1-\rho_{22}}{\widetilde{\mathsf{N}}^C}({\widetilde{\mathsf{M}}}^2{\widetilde{\mathsf{M}}}^4\cos^2\theta+{\widetilde{\mathsf{M}}}^1{\widetilde{\mathsf{M}}}^3\sin^2\theta),\nonumber\\
K_R&=\frac{1-\rho_{22}}{\widetilde{\mathsf{N}}^C}({\widetilde{\mathsf{M}}}^1{\widetilde{\mathsf{M}}}^3-{\widetilde{\mathsf{M}}}^2{\widetilde{\mathsf{M}}}^4)\sin\theta \cos\theta,\nonumber\\
L_R&=\frac{1-\rho_{22}}{2\widetilde{\mathsf{N}}^C}{\widetilde{\mathsf{M}}}^1{\widetilde{\mathsf{M}}}^4-\frac{\rho_{22}}{2}.
\end{align}
Analogously, the subscript or superscript $R$ represents the resonant case.
Correspondingly, the COA reads
\begin{equation}
C_\alpha^{R}=(1-h)\rho_{22}+h,
\end{equation}
where
\begin{align}
h=\frac{1}{\widetilde{\mathsf{N}}^C}[{\widetilde{\mathsf{M}}}^1{\widetilde{\mathsf{M}}}^4+2\sqrt{({\widetilde{\mathsf{M}}}^2{\widetilde{\mathsf{M}}}^4-{\widetilde{\mathsf{M}}}^1{\widetilde{\mathsf{M}}}^3)^2\sin^2\theta \cos^2\theta+{\widetilde{\mathsf{M}}}^1{\widetilde{\mathsf{M}}}^2{\widetilde{\mathsf{M}}}^3{\widetilde{\mathsf{M}}}^4}].
\end{align}
It is evident that the COA linearly depends on $\rho_{22}$. 

For the intuitive illustration, the COA and the HC are plotted in figure \ref{COA_heatcurrent_detuning}. It is shown in figure \ref{COA_heatcurrent_withTL_detuning} that the COA and the HC have the same trend with the temperature $T_L$, once the system is set. Even though the COA and the HC depend on the population $\rho_{22}$ in the resonant case, the same changing trend is well kept for the change of temperature $T_L$, which is given in figure \ref{withTL}.

\section{Discussion and Conclusion}
\label{sec:8}
Before the end, we’d like to mention that a number of rectifiers have been designed based on two-atom systems \cite{PhysRevE.95.022128,PhysRevE.99.042121,PhysRevResearch.2.033285,PhysRevE.104.054137}, but most of them are achieved by connecting two atoms to independent heat reservoirs. The CHR has been shown to have the advantage of enhancing the steady-state entanglement of the system \cite{liao2011quantum}, improving the efficiency of refrigerators \cite{manzano2019boosting} and the amplification of transistors \cite{e24010032}, etc. Based on two two-level atoms with CHRs, it is reported that quantum switch can be realized via two coupled superconducting qubits connecting RC oscillators \cite{karimi2017coupled}, and quantum diode with high rectification can be achieved by the same system connecting RLC oscillators \cite{PhysRevApplied.15.054050}. Comparably, we have designed a modulator utilizing the property of the dark state. 

In fact, the dark state is not a new concept, especially considering wide applications in quantum information and quantum optics, such as electromagnetic induced transparency (EIT) \cite{RevModPhys.77.633}, quantum nonreciprocity \cite{nefedkin2022dark} and so on. ‘Dark state’ does not evolve with time and is immune to decoherence and dissipation. It is attracting increasing interest in the potential applications in quantum thermal machines. For example, in reference \cite{PhysRevE.96.052126}, we use the dark state to block the heat current in a quantum refrigerator; In reference \cite{manzano2019boosting}, the dark state can double the performance of the refrigerator without any external resources; In reference \cite{e24010032}, we use the dark state to design quantum switch and in reference \cite{poulsen2022dark}, the heat diode is designed by using the property of blocking heat transfer.

In conclusion, we have studied the HCs of two coupled TLAs in CHRs using the BMS master equation and their analytic expressions. When the dissipation rates between the atoms and the heat reservoirs are not the same, we compared HCs in the cases of CHRs and IHRs of the same model.
When considering the influence of CHR with $g$ and $\omega_m$, we find that CHRs have a slight influence on HCs, indicating that the result of crossing dissipation on HCs is close to zero, but the CHR effect is negative because of the potentially present inverse heat currents. When we consider its variation with $\gamma_i$, it is found that CHR does not always inhibit HC but also could enhance it, which depends on the practical scenario such as $\gamma_i$, $\omega_m$, and $T_L$.
A significant result is that when the dissipation rates corresponding to the same eigenfrequency are equal, i.e., $\gamma_\alpha^{mn}(\omega_i)=\gamma_i$, the steady-state of the system is not unique and is a linear mixture of two states which include a 'dark state' contributing zero HC. So our TLAs model can act as a thermal modulator, in which we can continuously control the population $\rho_{22}$ by a single-mode laser and then continuously control HC. We also find that the variation tendencies of steady-state COA and the HC with temperature are consistent.

 This simple model shed new insight into relevant research in two aspects: CHR can create inverse heat currents which have no classical counterpart for macroscopic systems, and the steady-state thermal modulator can be effectively controlled by the initial states, which, via a dark state, determines the fractions of each component in steady state as well as the HCs.  

\invisiblesection{Acknowledgments}
\section*{Acknowledgments}
This work was supported by the National Natural Science
Foundation of China under Grant No.12175029, No.11775040, and No.12011530014.  

\appendices

\invisiblesection{Appendix}
\section*{Appendix A: Detail derivation of BMS master equation (\ref{MEq},\ref{MEq.(common)})}\label{AppendixA}
\renewcommand{\theequation}{A.\arabic{equation}}
\setcounter{equation}{0}  
Since both operators and quantum states are associated with time, we could shift the model to the interaction picture. So the interaction Hamiltonian is $\tilde{H}_{SE}(t)=A(t)\Upsilon_1(t)+B(t)\Upsilon_2(t)$, where
\begin{align}
A(t)&=\sum_j{f_{1j}^L(a_j^\dagger e^{i\omega_{Lj}t}+a_je^{-i\omega_{Lj}t})}+\sum_k{f_{1k}^R(b_k^\dagger e^{i\omega_{Rk}t}+b_ke^{-i\omega_{Rk}t})},\label{A(t)}\\
B(t)&=\sum_j{f_{2j}^R(a_j^\dagger e^{i\omega_{Lj}t}+a_je^{-i\omega_{Lj}t})}+\sum_k{f_{2k}^R(b_k^\dagger e^{i\omega_{Rk}t}+b_ke^{-i\omega_{Rk}t})}\label{B(t)}
\end{align}
are reservoir operators. It is clear that $A(t)=A^\dagger (t)$ and $B(t)=B^\dagger (t)$, and
\begin{align}
\Upsilon_1(t)=&\sum_{i=\pm}[V_{1}^\dagger (\omega_i)e^{i\omega_it}+V_{1}(\omega_i)e^{-i\omega_it}],\label{RNum(1)}\\
\Upsilon_2(t)=&\sum_{i=\pm}[V_{2}^\dagger (\omega_i)e^{i\omega_it}+V_{2}(\omega_i)e^{-i\omega_it}].\label{RNum(2)}
\end{align}

So the interaction Hamiltonian is
\begin{equation}
\tilde{H}_{SE}(t)=\sum_{i=\pm}[A(t)V_{1}^\dagger (\omega_i)+B(t)V_{2}^\dagger (\omega_i)]e^{i\omega_it}+\rm{H.c.}.
\label{tildeH_SE}
\end{equation}
Using the BMS approximation and Liouville equation $\dot{\rho}_{SE}(t)=-i[H,\rho_{SE}(t)]$, we get the master equation as
\begin{align}
\dot{\rho}(t)={\rm{Tr}}_E\int^\infty_0 d\tau[\tilde{H}_{SE}(t-\tau)\rho(t)\rho_B\tilde{H}_{SE}(t)-\tilde{H}_{SE}(t)\tilde{H}_{SE}(t-\tau)\rho(t)\rho_B]+\rm{H.c.},
\end{align}
where $\rho(t)=\rm{Tr}_E\rho_{SE}(t)$ describes the reduced density matrix of the system. If we substitute equation (\ref{tildeH_SE}) into the above equation, we get
\begin{align}
\nonumber
\dot{\rho}(t)&=\sum_{i=\pm}\sum_{m,n=1}^2[V_{m}^\dagger (\omega_i)\rho(t) V_{n}(\omega_i)-V_{n}(\omega_i)V_{m}^\dagger (\omega_i)\rho(t)]\int_0^\infty d\tau\langle N(\tau)M(0)\rangle e^{-i\omega_i\tau}\\
&+[V_{m}(\omega_i)\rho(t) V_{n}^\dagger (\omega_i)-V_{n}^\dagger (\omega_i)V_{m}(\omega_i)\rho(t)]\int_0^\infty d\tau\langle N(0)M(-\tau)\rangle e^{i\omega_i\tau}+\rm{H.c.}.
\label{intrho}
\end{align}
Here, we regard $i$ as the index of the atomic eigenfrequencies, $m$ and $n$ denote the indices of the atoms, i.e., $1$ represents the first atom and $2$ is on behalf of the second one, and $M$ (or $N$) denotes the reservoir function $A(t)$ (or $B(t)$) connect to $1_{th}$ (or $2_{th}$) atom, respectively. Using the properties of the reservoir correlation functions $\langle N(t)M(t-\tau)\rangle\equiv{\rm{Tr}}_E[N(t)M(t-\tau)\rho_E]=\langle N(\tau)M(0)\rangle=\langle N(0)M(-\tau)\rangle$, we can get equation (\ref{intrho}). Next, we calculate the one-sided Fourier transforms of the heat reservoir correlation function. Let's pick one of them as an example,
\begin{align}\label{example}
\nonumber
&\int_0^\infty d\tau \langle A(0)B(-\tau)e^{-i\omega_-\tau}\\
\nonumber
=&\sum_j[f_{1j}^Lf_{2j}^L(\langle a_j^\dagger a_j\rangle\int_0^\infty d\tau e^{i(\omega_{Lj}-\omega_-)\tau}+\langle a_ja_j^\dagger \rangle\int_0^\infty d\tau e^{-i(\omega_{Lj}+\omega_-)\tau})]\\
&+\sum_k[f_{1k}^Rf_{2k}^R(\langle b_k^\dagger b_k\rangle\int_0^\infty d\tau e^{i(\omega_{Rk}-\omega_-)\tau}+\langle b_kb_k^\dagger \rangle\int_0^\infty d\tau e^{-i(\omega_{Rk}+\omega_-)\tau})].
\end{align}
The formula
\begin{equation}
\int_0^\infty d\tau e^{-i\omega\tau}=\pi\delta(\tau)-i{\rm{P}}\frac{1}{\omega},
\end{equation}
where $\rm{P}$ represents the Cauchy principal value and we will not consider the imaginary part in the following calculation since it will eventually be summed up as the Lamb shift term. Since heat reservoirs are composed of infinite dimensional harmonic oscillators, we have
\begin{alignat}{2}
\langle a_j^\dagger a_j\rangle &=\bar{n}_L(\omega_{Lj}), &\quad \langle a_ja_j^\dagger \rangle &= \bar{n}_L(\omega_{Lj})+1,\\
\langle b_k^\dagger b_k\rangle &=\bar{n}_R(\omega_{Rk}), &\quad \langle b_kb_k^\dagger \rangle &= \bar{n}_R(\omega_{Rk})+1.
\end{alignat}
Thus equation (\ref{example}) follows
\begin{align}
\int_0^\infty d\tau \langle A(0)B(-\tau)e^{-i\omega_-\tau}=\gamma_L^{12}(\omega_-)\bar{n}_L(\omega_-)+\gamma_R^{12}(\omega_-)\bar{n}_R(\omega_-),\label{int}
\end{align}
where $\gamma_{L/R}^{mn}(\omega_i)=\pi f_{nj/k}^{L/R}(\omega_i)f_{mj/k}^{L/R}(\omega_i)$ is the dissipation factor and represents the dissipation strength between the atom and the heat reservoirs. We take $J_\alpha^{mn}(\pm\omega_i)=\gamma_\alpha^{mn}(\omega_i)[\pm\bar{n}_\alpha(\pm\omega_i)]$ as the spectral density, and $\bar{n}_\alpha(\omega_i)=\frac{1}{e^{\frac{\omega_i}{T_{\alpha}}}-1}$ is the average number of photons corresponding to the frequency $\omega_i$ of the $\alpha_{th}$ heat reservoir.
After the detailed calculation, the master equation is
\begin{align}
\nonumber
\dot{\rho}(t)=&\sum_{\alpha=L,R}\sum_{i=\pm}\sum_{m,n=1}^2 J^{mn}_\alpha(-\omega_i)[2V_{n}(\omega_i)\rho V^\dagger _{m}(\omega_i)-V^\dagger _{m}(\omega_i)V_{n}(\omega_i)\rho-\rho V^\dagger _{m}(\omega_i)V_{n}(\omega_i)]\\
&+J_\alpha^{mn}(\omega_i)[2V^\dagger _{n}(\omega_i)\rho V_{m}(\omega_i)-V_{m}(\omega_i)V^\dagger _{n}(\omega_i)\rho-\rho V_{m}(\omega_i)V^\dagger _{n}(\omega_i)].
\end{align}

\section*{Appendix B: Detail derivation of BMS master equation (\ref{MEq},\ref{MEq.(independent)})}\label{AppendixB}
\renewcommand{\theequation}{B.\arabic{equation}}
\setcounter{equation}{0}  

\begin{figure}
	\centering
		\hspace{-0.8cm}
	\subfigbottomskip=2pt
	\subfigure[ ]{
		\includegraphics[width=5cm]{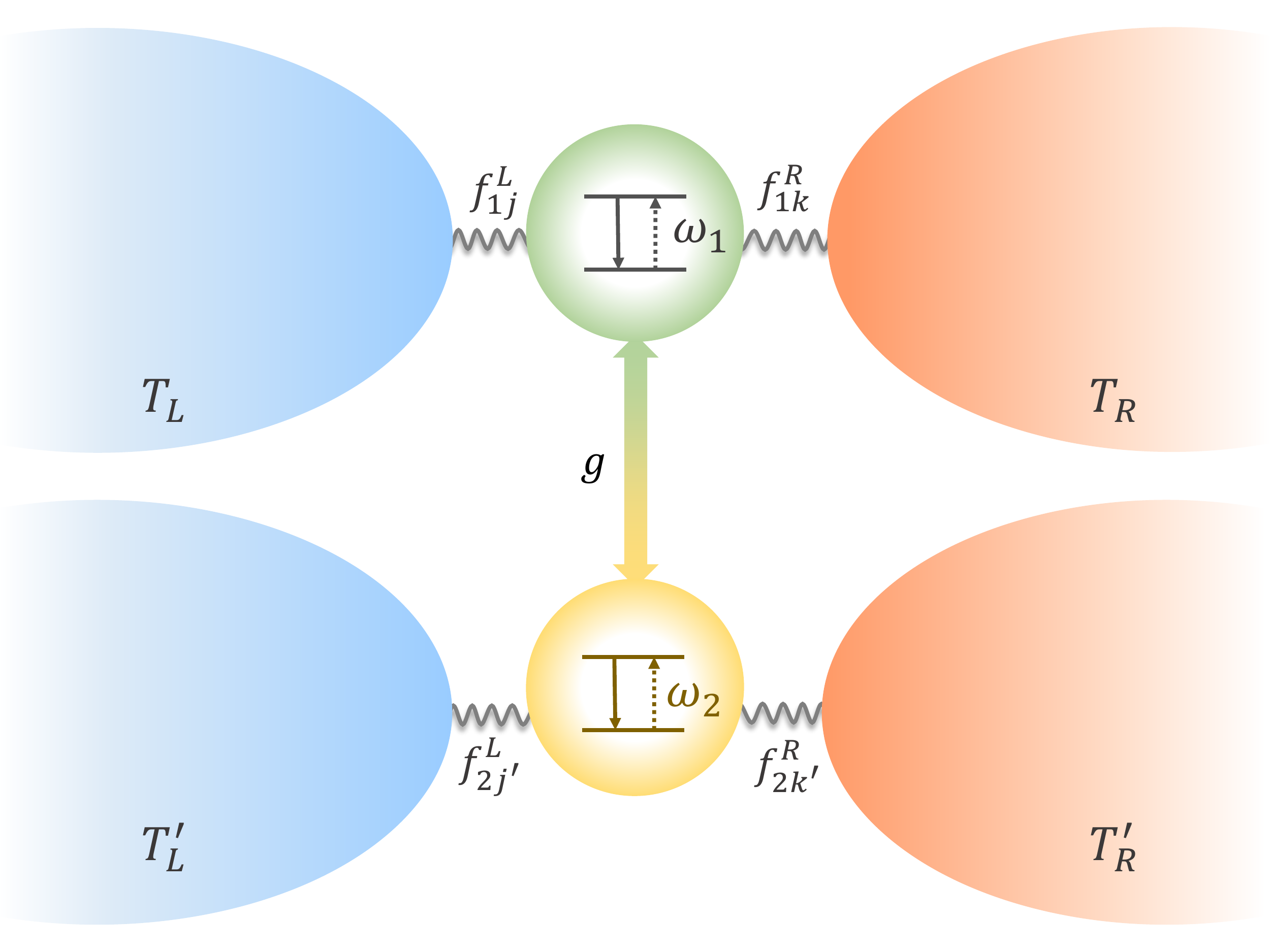}
		\label{model_independen}}
	\subfigure[ ]{
		\includegraphics[width=5cm]{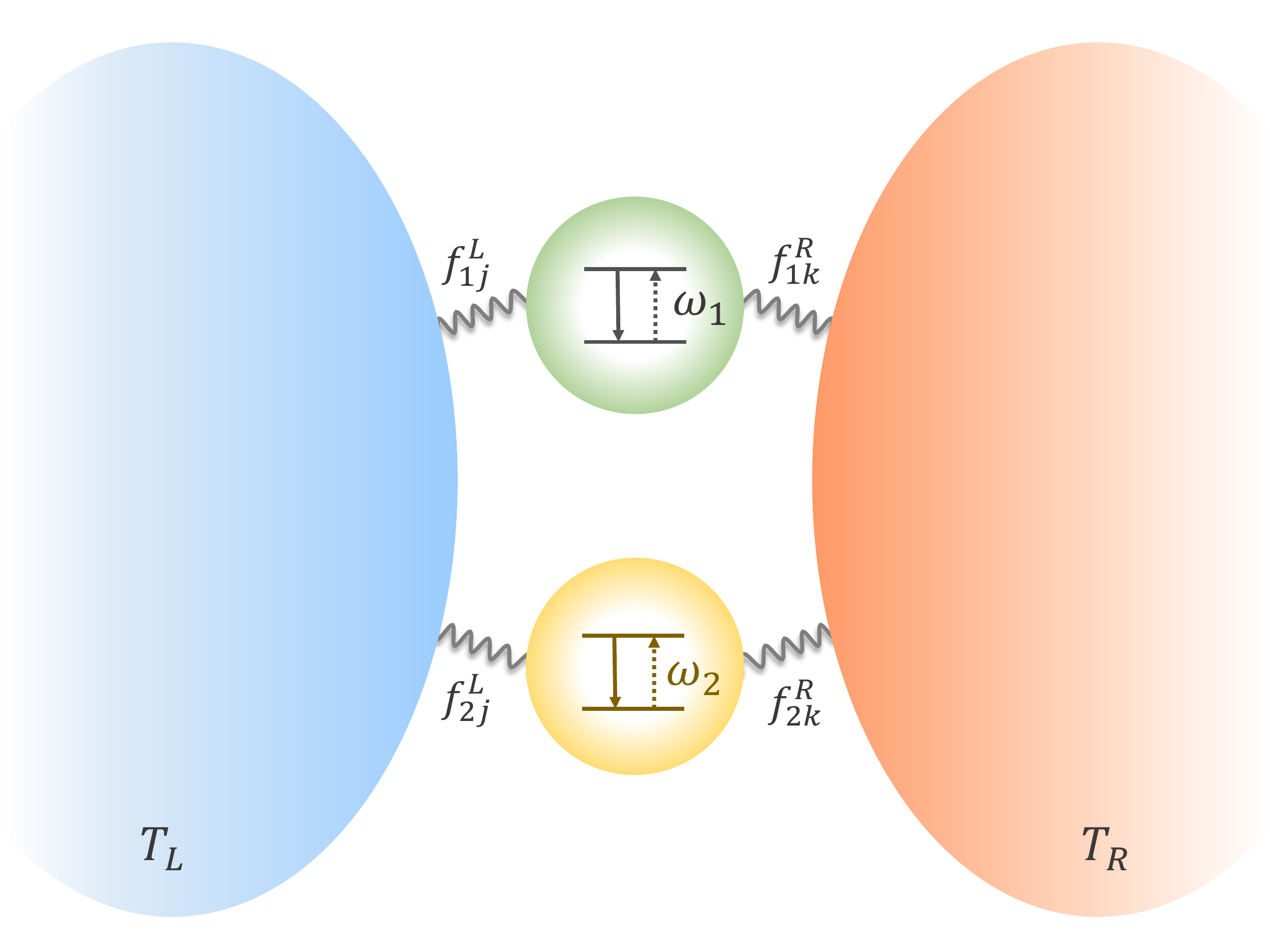}
		\label{model_commong0}}
	\subfigure[ ]{
		\includegraphics[width=5cm]{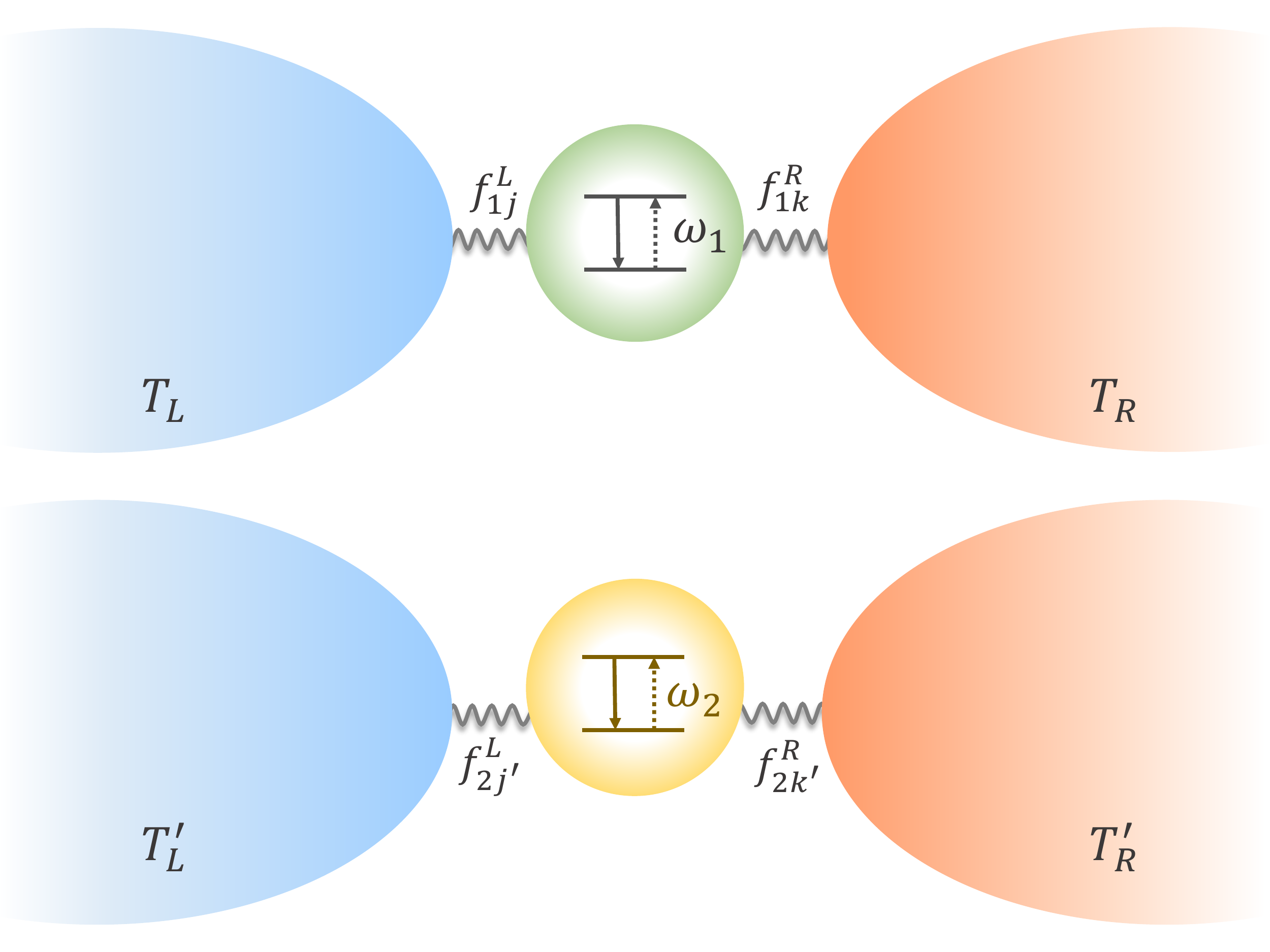}\label{model_independentg0}}
	\caption{The different cases of interatomic coupling and atomic coupling to the heat reservoirs. (a) denotes two coupled two-level atoms with the coupling strength $g$ are connected to four IHRs which the temperatures, in turn, are $T_L$, $T_R$, $T_L'$ and $T_R'$, the coupling factors are $f_{1j}^L$, $f_{1k}^R$, $f_{2j'}^L$ and $f_{2k'}^R$, the numbers $j, k, j'$ and $k'$ denote the modes of the corresponding heat reservoirs. (b) denotes two uncoupled two-level atoms, i.e., $g=0$, are connected to two CHRs which the temperatures are $T_{L}$ and $T_{R}$, the coupling factors are $f_{1j}^L$, $f_{1k}^R$, $f_{2j}^L$ and $f_{2k}^R$, the numbers $j$ and $k$ denote the modes of the corresponding heat reservoirs. (c) denotes two uncoupled $g=0$ two-level atoms are connected to two heat reservoirs independently as (a).}
\end{figure}

Let's consider the case of IHRs as shown in figure~\ref{model_independen}.
The difference in the Hamiltonian between IHR and CHR lies in $H_E$ and $H_{SE}$. Their specific forms are given as
\begin{align}
H'_{E}=\sum_j \omega_{Lj}a_j^\dagger a_j+\sum_k \omega_{Rk}b_k^\dagger b_k+\sum_{j'}\omega_{L'j'}{a'}_{j'}^\dagger {a'}_{j'}+\sum_{k'}\omega_{R'k'}{b'}_{k'}^\dagger {b'}_{k'}
\end{align}
and 
\begin{align}
\nonumber
H'_{SE}=&[\sum_j f_{1j}^L(a_j^\dagger +a_j)+\sum_k f_{1k}^R(b_k^\dagger +b_k)] \sigma_1^x\\
&+[\sum_{j'} f_{2j'}^L({a'}_{j'}^\dagger +{a'}_{j'})+\sum_{k'}^R f_{2k'}({b'}_{k'}^\dagger +{b'}_{k'})] \sigma_2^x.
\end{align}

Since the atomic system is unchanged, the diagonal $H_S$, the eigenoperators, and corresponding eigenfrequencies are consistent with the case of CHRs. In the interaction picture, the interaction Hamiltonian is
\begin{equation}
\tilde{H}'_{SE}(t)=A(t)\Upsilon_1(t)+B'(t)\Upsilon_2(t),
\end{equation}
where
\begin{align}
B'(t)=\sum_{j'}{f_{2j'}^L({a'}_{j'}^\dagger e^{i\omega_{L'j'}t}+{a'}_{j'}e^{-i\omega_{L'j'}t})}+\sum_{k'}{f_{2k'}^R({b'}_{k'}^\dagger e^{i\omega_{R'k'}t}+{b'}_{k'}e^{-i\omega_{R'k'}t})}.
\end{align}
$B'(t)=B'^\dagger (t)$ holds and $A(t)$, $\Upsilon_1(t)$ and $\Upsilon_2(t)$ are the same as equations (\ref{A(t)},\ref{RNum(1)},\ref{RNum(2)}).
So the interaction Hamiltonian is
\begin{equation}
\tilde{H}'_{SE}(t)=\sum_{i=\pm}[A(t)V_1^\dagger (\omega_i)+B'(t)V_2^\dagger (\omega_i)]e^{i\omega_it}+\rm{H.c.}.
\end{equation} Following the steps in appendix \hyperref[AppendixA]{A}, since the average values of the generation and annihilation operators of different heat reservoirs are all zero, for example, $\langle a^\dagger a'\rangle=0$, equation (\ref{int}) becomes $\int_0^\infty d\tau \langle A(0)B'(-\tau)e^{-i\omega_-\tau}=0$. Hence only the direct coupling dissipative term exists. We can obtain the master equation in the interaction picture the same as equation (\ref{MEq}), i.e.,
\begin{equation}
\mathcal{L}_{\alpha}(\rho)=\mathcal{L}_{\alpha}^{11}(\rho)+\mathcal{L}_{\alpha}^{22}(\rho).
\end{equation}

Using the same method, we can directly calculate HCs without the coupling between atoms as shown in figure \ref{model_commong0}. The Hamiltonian of the system is
\begin{equation}
H_S=\frac{\omega_1}{2}\sigma_1^z+\frac{\omega_2}{2}\sigma_2^z=\sum_{l'=1}^4\lambda_{l'}\vert l'\rangle\langle l'\vert,
\end{equation}
where $[\lambda_{1'},\lambda_{2'},\lambda_{3'},\lambda_{4'}]=[-\omega_s,-\omega_d,\omega_d,\omega_s]$, and the corresponding eigenstates are $\vert1'\rangle=\vert\downarrow\downarrow\rangle$, $\vert2'\rangle=\vert\downarrow\uparrow\rangle$, $\vert3'\rangle=\vert\uparrow\downarrow\rangle$ and $\vert4'\rangle=\vert\uparrow\uparrow\rangle$. $H_E$ and $H_{SE}$ are the same as equation (\ref{H_E}) and equation (\ref{H_(SE)}). The eigenoperators are
\begin{align}
V_1(\omega_1)&=\vert1'\rangle\langle 3'\vert+\vert2'\rangle\langle4'\vert=\sigma_1^-\otimes\mathbbm{1}_2,\label{V1}\\
V_2(\omega_2)&=\vert1'\rangle\langle 2'\vert+\vert3'\rangle\langle4'\vert=\mathbbm{1}_1\otimes\sigma^-_2,\label{V2}
\end{align}
where $\sigma^-_m=\vert\downarrow\rangle_m\langle\uparrow\vert$ and the eigenfrequencies are their natural frequencies. Then we can get the master equation as equation (\ref{MEq}) with
\begin{align}
\nonumber
\mathcal{L}_{\alpha}(\rho)
=&\sum_{m=1}^2J_{\alpha}^{mm}(-\omega_m)(2\sigma_m^-\rho \sigma_m^+-\sigma_m^+\sigma_m^-\rho-\rho \sigma_m^+\sigma_m^-)\\
\nonumber
&+J_{\alpha}^{mm}(\omega_m)(2\sigma_m^+\rho\sigma_m^--\sigma_m^-\sigma_m^+\rho-\rho \sigma_m^-\sigma_m^+),\\
\end{align}
where $J_{\alpha}^{mm}(\pm\omega_m)$ has the same definition as before.

\textit{When the two TLAs are in resonance}, the crossing-coupling term occurs and the dissipation becomes
\begin{align}
\nonumber
\mathcal{L}_{\alpha}(\rho)
=&\sum_{m,n=1}^2J_{\alpha}^{mn}(-\omega)(2\sigma^-_n\rho\sigma^+_m-\sigma^+_m\sigma^-_n\rho-\rho\sigma^+_m\sigma^-_n)\\
&+J_{\alpha}^{mn}(\omega)(2\sigma^+_n\rho\sigma^-_m-\sigma^-_m\sigma^+_n\rho-\rho\sigma^-_m\sigma^+_n).\label{dissipatorgeq0}
\end{align}

As mentioned in section \ref{sec:2}, when atoms are connected to common baths, dissipations corresponding to the same eigenfrequency $\mathcal{L}_\alpha^{mm}(\omega_i)$ and $\mathcal{L}_\alpha^{nn}(\omega_i)$ must lead to cross term $\mathcal{L}_\alpha^{mn}(\omega_i)$ no matter whether there is coupling between atoms. The above equation amply illustrates this point. 
Note that the existence of the cross dissipation without interatomic coupling doesn't contradict with the Born-Markov approximation which emphasizes that the state of the environment is not affected by the interaction between atoms and reservoirs. No matter how the system interacts with the reservoirs, the states of the reservoirs aren't changed, but their effect on the system of interest should depend on the concrete interaction between them. In the current case, the CHRs are just shown by the cross dissipation.

\textit{We can also calculate the case of uncoupled atoms' contact with IHRs}. Since each atom is connected to two reservoirs separately and no other connection between the atoms as shown in figure \ref{model_independentg0}, we only consider the top half for convenience, that is, one atom is connected to two heat reservoirs, so the total Hamiltonian is
\begin{align}
H=\frac{\omega}{2}{\sigma}^z+\sum_{j}\omega_{Lj}a^\dagger _ja_j+\sum_{k}\omega_{Rk}b^\dagger _kb_k+[\sum_{j}f_j^L(a_j^\dagger +a_j)+\sum_{k}f_k^R(b_k^\dagger +b_k)]{\sigma}^x.
\end{align}
There is only one eigenoperator $V(\omega)=\sigma^-$ and the corresponding eigenfrequency is $\omega$. In this case, the master equation also can be written as equation (\ref{MEq}), but the dissipation is
\begin{align}
\nonumber
\mathcal{L}_\alpha(\rho)
=&J_\alpha(-\omega)(2\sigma^-\rho\sigma^+-\sigma^+\sigma^-\rho-\rho\sigma^+\sigma^-)\\
\nonumber
&+J_\alpha(\omega)(2\sigma^+\rho\sigma^--\sigma^-\sigma^+\rho-\rho\sigma^-\sigma^+).\\
\end{align}
The dissipation of the lower part is the same as the above formula, except that the eigenfrequency is its natural frequency.

\section*{Appendix C: Steady state with all dissipation rates exactly the same}
\label{AppendixC}
\renewcommand{\theequation}{C.\arabic{equation}}
\setcounter{equation}{0} 
In section \ref{sec:3}, we specifically derived the steady state and HC between the system and IHRs or CHRs. When we consider the special case of $\gamma_\alpha^{mn}(\omega_i)=\gamma_i$, by simplification, we get that the steady state in both environments can be expressed as $\vert\varrho\rangle=\frac{1}{N}[\varrho_{11},\varrho_{22},\varrho_{33},\varrho_{44}]^T$ and the specific form as
\begin{align}
\varrho_{11}&=[n(\omega_-)+2][n(\omega_+)+2]\label{rho1n},\\
\varrho_{22}&=n(\omega_-)[n(\omega_+)+2]\label{rho2n},\\
\varrho_{33}&=[n(\omega_-)+2]n(\omega_+)\label{rho3n},\\
\varrho_{44}&=n(\omega_-)n(\omega_+)\label{rho4n},
\end{align}
where $n(\omega_i)=n_R(\omega_i)+n_L(\omega_i)$ and the normalization coefficient $N=\sum_{p=1}^4\varrho_{pp}$.
It is obvious that $\varrho_{11}>\varrho_{22}>\varrho_{33}>\varrho_{44}$. 
Since the same steady state can be obtained no matter whether the system is connected to IHR or CHR, the effect of CHR is only taken by CHC, i.e., equation (\ref{DetuningQc}), which can be found by comparing equations (\ref{heat_IHR},\ref{DetuningQd}). 
Now, CHC can be rewritten as
\begin{align}
\nonumber
\dot{Q}_L^c=&\frac{8}{N}\{\cos\theta_s\sin\theta_s[\gamma_-\omega_-u(\omega_-)+\gamma_+\omega_+u(\omega_+)]+\cos\theta_d\sin\theta_d[\gamma_-\omega_-u(\omega_-)-\gamma_+\omega_+u(\omega_+)]\}\\
\nonumber
=&\frac{4}{N}[\gamma_-\omega_-(\cos\theta_s\sin\theta_s+\cos\theta_d\sin\theta_d)(\coth\frac{\omega_-}{2T_R}-\coth\frac{\omega_-}{2T_L})\\
&+\gamma_+\omega_+(\cos\theta_s\sin\theta_s-\cos\theta_d\sin\theta_d)(\coth\frac{\omega_+}{2T_R}-\coth\frac{\omega_+}{2T_L})],\label{QLcross}
\end{align}
where the forms of $\sin\theta_\nu$ and $\cos\theta_\nu$ are found in section \ref{sec:2} and they are always positive, $u(\omega_i)=n_R(\omega_i)-n_L(\omega_i)=\frac{1}{2}(\coth\frac{\omega_i}{2T_R}-\coth\frac{\omega_i}{2T_L})<0$ for $T_R<T_L$. Through the concrete numerical verification ('NMaximize' in Mathematica 12.3), we find that equation (\ref{QLcross}) is always less than zero under the condition $\gamma_-=\gamma_+=\gamma$, that is, $\dot{Q}_L^c<0$ always holds. However, when $\gamma_-\neq\gamma_+$, $\dot{Q}_L^c>0$ is possible.  

\section*{Appendix D: The evolution with driving}
\label{AppendixD}
\renewcommand{\theequation}{D.\arabic{equation}}
\setcounter{equation}{0}  
Since the population of the second energy level $\rho_{22}$ plays a decisive role in HC in the $H_S$ representation, we can use it to design a heat modulator. In this part, we derive in detail the population of each energy level at any time. For simplicity, we do not consider the first and fourth energy levels, so the system is reduced to a two-level system. Here, we express the electric dipole interaction Hamiltonian between single-mode electromagnetic field $\vec{E}(t)=\vec{\epsilon}\cos(\nu_{23}t)$ and the system as
\begin{align}
\nonumber
H_{I}(t)=&-\vec{r}\cdot\vec{E}(t)\\
\nonumber
=&-(\vert3\rangle\langle3\vert+\vert2\rangle\langle2\vert)\vec{r}(\vert3\rangle\langle3\vert+\vert2\rangle\langle2\vert)\cdot\vec{E}(t)\\
\nonumber
=&-(\vec{r}_{32}\vert3\rangle\langle2\vert+\vec{r}_{23}\vert2\rangle\langle3\vert)\cdot\vec{\epsilon}\cos(\nu_{23}t)\\
=&\Omega_R(\vert3\rangle\langle2\vert+\vert2\rangle\langle3\vert)\cos(\nu_{23}t),
\end{align}
where $\Omega_R=-\vec{r}_{23}\cdot\vec{\epsilon}$ is the Rabi frequency and $\vec{r}_{23}=\langle2\vert\vec{r}\vert3\rangle=(\vec{r}_{32})^*$ is the matrix element of the electric dipole moment. The Hamiltonian of the whole system is
\begin{equation}
H=w_3\vert3\rangle\langle3\vert+w_2\vert2\rangle\langle2\vert+\Omega_R(\vert3\rangle\langle2\vert+\vert2\rangle\langle3\vert)\cos\nu_{23}t
\end{equation}
and we define $\vec{r}_{23}=\vert\vec{r}_{23}\vert e^{i\phi}$ and the phase $\phi=0$, $\vert\vec{r}_{23}\vert=\vert\vec{r}_{32}\vert$.
Thus we can obtain the state $\vert\psi(t)\rangle$ at any time by solving the Schr$\ddot{\rm{o}}$dinger equation
\begin{equation}
\frac{d}{dt}\vert\psi(t)\rangle=-iH\vert\psi(t)\rangle,
\end{equation}
where $\vert\psi(t)\rangle=\left(
\begin{smallmatrix}
c(t)\\b(t)
\end{smallmatrix}
\right)$. One can get the differential equations
\begin{align}
\dot{c}(t)&=-iw_3c(t)-i\Omega_R\cos(\nu_{23}t)b(t),\\
\dot{b}(t)&=-iw_2b(t)-i\Omega_R\cos(\nu_{23}t)c(t).
\end{align}
Define
\begin{align}
C(t)&=c(t)e^{iw_3t},\\
B(t)&=b(t)e^{iw_2t},
\end{align}
we can obtain
\begin{align}
\dot{C}(t)&=-i\frac{\Omega_R}{2}B(t)e^{i\Delta t},\label{dotC}\\
\dot{B}(t)&=-i\frac{\Omega_R}{2}C(t)e^{-i\Delta t},
\end{align}
where $\Delta=w_{32}-\nu_{23}$, $w_{32}=w_3-w_2$. For convenience, we make the system resonate with this single-mode laser as $\Delta=0$. Take the derivative of equation (\ref{dotC}) with respect to time, we get
\begin{equation}
\ddot{C}(t)+\frac{\Omega_R^2}{4}C(t)=0.
\end{equation}
If we set the initial state to be $\vert\psi(0)\rangle=\left(
\begin{smallmatrix}
C(0)\\B(0)
\end{smallmatrix}
\right)$, the solution $\vert\psi(t)\rangle$ is
\begin{equation}
\vert\psi(t)\rangle=\left(
\begin{matrix}
[\cos(\frac{\Omega_R t}{2})C(0)-i\sin(\frac{\Omega_R t}{2})B(0)]e^{iw_3 t}\\
[\cos(\frac{\Omega_R t}{2})B(0)-i\sin(\frac{\Omega_R t}{2})C(0)]e^{iw_2 t}
\end{matrix}
\right).
\end{equation}
The populations can be given as 
\begin{align}
\rho_{33}(t)&=\frac{A^-}{2}\cos(\Omega_R t)+\frac{A^+}{2},\\
\rho_{22}(t)&=-\frac{A^-}{2}\cos(\Omega_R t)+\frac{A^+}{2},
\end{align}
where $A^\pm=|C(0)|^2 \pm |B(0)|^2$.

\section*{References}

\bibliographystyle{iopart-num}
\bibliography{Common_bib} 

\providecommand{\noopsort}[1]{}\providecommand{\singleletter}[1]{#1}%
\providecommand{\newblock}{}
\begin{thebibliography}{10}
\expandafter\ifx\csname url\endcsname\relax
  \def\url#1{{\tt #1}}\fi
\expandafter\ifx\csname urlprefix\endcsname\relax\def\urlprefix{URL }\fi
\providecommand{\eprint}[2][]{\url{#2}}

\bibitem{landsberg1956foundations}
Landsberg P~T 1956 {\em Rev. Mod. Phys.\/} {\bf 28}(4) 363--392
  \urlprefix\url{https://link.aps.org/doi/10.1103/RevModPhys.28.363}

\bibitem{parrondo2015thermodynamics}
Parrondo J~M~R, Horowitz J~M and Sagawa T 2015 {\em Nat. Phys.\/} {\bf 11}
  131--139 \urlprefix\url{https://doi.org/10.1038/NPHYS3230}

\bibitem{sakurai1995modern}
Sakurai J~J and Commins E~D 1995 Modern quantum mechanics, revised edition

\bibitem{ballentine2014quantum}
Ballentine L~E 2014 {\em Quantum mechanics: a modern development\/} (World
  Scientific Publishing Company)

\bibitem{zohar1990quantum}
Zohar D and Marshall I~N 1990 {\em The quantum self: Human nature and
  consciousness defined by the new physics.\/} (William Morrow \& Co)
  \urlprefix\url{https://psycnet.apa.org/record/1991-97708-000}

\bibitem{lemaitre1931beginning}
Lema{\^\i}tre G 1931 {\em Nature\/} {\bf 127} 706--706
  \urlprefix\url{https://doi.org/10.1038/127706b0}

\bibitem{cao2021quantum}
Cao X, Wang C, Zheng H and He D 2021 {\em Phys. Rev. B\/} {\bf 103}(7) 075407
  \urlprefix\url{https://link.aps.org/doi/10.1103/PhysRevB.103.075407}

\bibitem{hewgill2021quantum}
Hewgill A, De~Chiara G and Imparato A 2021 {\em Phys. Rev. Research\/} {\bf
  3}(1) 013165
  \urlprefix\url{https://link.aps.org/doi/10.1103/PhysRevResearch.3.013165}

\bibitem{quan2007quantum}
Quan H~T, Liu Y~x, Sun C~P and Nori F 2007 {\em Phys. Rev. E\/} {\bf 76}(3)
  031105 \urlprefix\url{https://link.aps.org/doi/10.1103/PhysRevE.76.031105}

\bibitem{liu2021periodically}
Liu J, Jung K~A and Segal D 2021 {\em Phys. Rev. Lett.\/} {\bf 127}(20) 200602
  \urlprefix\url{https://link.aps.org/doi/10.1103/PhysRevLett.127.200602}

\bibitem{feldmann2000performance}
Feldmann T and Kosloff R 2000 {\em Phys. Rev. E\/} {\bf 61}(5) 4774--4790
  \urlprefix\url{https://link.aps.org/doi/10.1103/PhysRevE.61.4774}

\bibitem{palao2001quantum}
Palao J~P, Kosloff R and Gordon J~M 2001 {\em Phys. Rev. E\/} {\bf 64}(5)
  056130 \urlprefix\url{https://link.aps.org/doi/10.1103/PhysRevE.64.056130}

\bibitem{ccakmak2020quantum}
{\c{C}}akmak S and Altintas F 2020 {\em Quantum Inf. Process.\/} {\bf 19} 1--15
  \urlprefix\url{https://doi.org/10.1007/s11128-020-02746-x}

\bibitem{humphrey2002reversible}
Humphrey T~E, Newbury R, Taylor R~P and Linke H 2002 {\em Phys. Rev. Lett.\/}
  {\bf 89}(11) 116801
  \urlprefix\url{https://link.aps.org/doi/10.1103/PhysRevLett.89.116801}

\bibitem{lu2020brownian}
Lu J, Wang R, Wang C and Jiang J~H 2020 {\em Phys. Rev. B\/} {\bf 102}(12)
  125405 \urlprefix\url{https://link.aps.org/doi/10.1103/PhysRevB.102.125405}

\bibitem{wang2002experimental}
Wang G~M, Sevick E~M, Mittag E, Searles D~J and Evans D~J 2002 {\em Phys. Rev.
  Lett.\/} {\bf 89}(5) 050601
  \urlprefix\url{https://link.aps.org/doi/10.1103/PhysRevLett.89.050601}

\bibitem{kosloff2013quantum}
Kosloff R 2013 {\em Entropy\/} {\bf 15} 2100--2128
  \urlprefix\url{https://doi.org/10.3390/e15062100}

\bibitem{levy2014local}
Levy A and Kosloff R 2014 {\em Europhys. Lett.\/} {\bf 107} 20004
  \urlprefix\url{https://iopscience.iop.org/article/10.1209/0295-5075/107/20004/meta}

\bibitem{lieb2000fresh}
Lieb E~H and Yngvason J 2000 A fresh look at entropy and the second law of
  thermodynamics {\em Statistical Mechanics\/} (Springer) pp 365--370
  \urlprefix\url{https://doi.org/10.1007/978-3-662-10018-9_20}

\bibitem{quan2006maxwell}
Quan H~T, Wang Y~D, Liu Y~x, Sun C~P and Nori F 2006 {\em Phys. Rev. Lett.\/}
  {\bf 97}(18) 180402
  \urlprefix\url{https://link.aps.org/doi/10.1103/PhysRevLett.97.180402}

\bibitem{strasberg2013thermodynamics}
Strasberg P, Schaller G, Brandes T and Esposito M 2013 {\em Phys. Rev. Lett.\/}
  {\bf 110}(4) 040601
  \urlprefix\url{https://link.aps.org/doi/10.1103/PhysRevLett.110.040601}

\bibitem{cottet2017observing}
Cottet N, Jezouin S, Bretheau L, Campagne-Ibarcq P, Ficheux Q, Anders J,
  Auff{\`e}ves A, Azouit R, Rouchon P and Huard B 2017 {\em Proc. Natl. Acad.
  Sci. U.S.A.\/} {\bf 114} 7561--7564
  \urlprefix\url{https://doi.org/10.1073/pnas.1704827114}

\bibitem{binder2019thermodynamics}
Binder F, Correa L~A, Gogolin C, Anders J and Adesso G 2019 {\em Thermodynamics
  in the quantum regime: fundamental aspects and new directions\/} vol 195
  (Springer)

\bibitem{millen2016perspective}
Millen J and Xuereb A 2016 {\em New J. Phys.\/} {\bf 18} 011002
  \urlprefix\url{https://iopscience.iop.org/article/10.1088/1367-2630/18/1/011002/meta}

\bibitem{vinjanampathy2016quantum}
Vinjanampathy S and Anders J 2016 {\em Contemp. Phys.\/} {\bf 57} 545--579
  \urlprefix\url{https://doi.org/10.1080/00107514.2016.1201896}

\bibitem{ThreeLevelMasers}
Scovil H~E~D and Schulz-DuBois E~O 1959 {\em Phys. Rev. Lett.\/} {\bf 2}(6)
  262--263 \urlprefix\url{https://link.aps.org/doi/10.1103/PhysRevLett.2.262}

\bibitem{yu2014re}
Yu C~s and Zhu Q~y 2014 {\em Phys. Rev. E\/} {\bf 90}(5) 052142
  \urlprefix\url{https://link.aps.org/doi/10.1103/PhysRevE.90.052142}

\bibitem{venturelli2013minimal}
Venturelli D, Fazio R and Giovannetti V 2013 {\em Phys. Rev. Lett.\/} {\bf
  110}(25) 256801
  \urlprefix\url{https://link.aps.org/doi/10.1103/PhysRevLett.110.256801}

\bibitem{hofer2016autonomous}
Hofer P~P, Perarnau-Llobet M, Brask J~B, Silva R, Huber M and Brunner N 2016
  {\em Phys. Rev. B\/} {\bf 94}(23) 235420
  \urlprefix\url{https://link.aps.org/doi/10.1103/PhysRevB.94.235420}

\bibitem{yu2019quantum}
Yu C~S, Guo B~Q and Liu T 2019 {\em Opt. Express\/} {\bf 27} 6863--6877
  \urlprefix\url{https://doi.org/10.1364/OE.27.006863}

\bibitem{hofer2017quantum}
Hofer P~P, Brask J~B, Perarnau-Llobet M and Brunner N 2017 {\em Phys. Rev.
  Lett.\/} {\bf 119}(9) 090603
  \urlprefix\url{https://link.aps.org/doi/10.1103/PhysRevLett.119.090603}

\bibitem{yang2019thermal}
Yang J, Elouard C, Splettstoesser J, Sothmann B, S\'anchez R and Jordan A~N
  2019 {\em Phys. Rev. B\/} {\bf 100}(4) 045418
  \urlprefix\url{https://link.aps.org/doi/10.1103/PhysRevB.100.045418}

\bibitem{wang2019thermal}
Wang C, Xu D, Liu H and Gao X 2019 {\em Phys. Rev. E\/} {\bf 99}(4) 042102
  \urlprefix\url{https://link.aps.org/doi/10.1103/PhysRevE.99.042102}

\bibitem{karimi2017coupled}
Karimi B, Pekola J~P, Campisi M and Fazio R 2017 {\em Quantum Sci. Technol.\/}
  {\bf 2} 044007 \urlprefix\url{https://doi.org/10.1088/2058-9565/aa8330}

\bibitem{farsani2019quantum}
Farsani M~J and Fazio R 2019 {\em Phys. Lett. A\/} {\bf 383} 1722--1727
  \urlprefix\url{https://doi.org/10.1016/j.physleta.2019.02.045}

\bibitem{lashkaryov1941investigations}
Lashkaryov V 1941 {\em Izv. Akad. Nauk SSSR, Ser. Fiz\/} {\bf 5} 442--446

\bibitem{li2004thermal}
Li B, Wang L and Casati G 2004 {\em Phys. Rev. Lett.\/} {\bf 93}(18) 184301
  \urlprefix\url{https://link.aps.org/doi/10.1103/PhysRevLett.93.184301}

\bibitem{werlang2014optimal}
Werlang T, Marchiori M~A, Cornelio M~F and Valente D 2014 {\em Phys. Rev. E\/}
  {\bf 89}(6) 062109
  \urlprefix\url{https://link.aps.org/doi/10.1103/PhysRevE.89.062109}

\bibitem{PhysRevE.99.042121}
Karg\ifmmode \imath \else~\i \fi{} C, Naseem M~T, Opatrn\'y T~c~v,
  M\"ustecapl\ifmmode \imath \else \i \fi{}o\ifmmode~\breve{g}\else
  \u{g}\fi{}lu O~E and Kurizki G 2019 {\em Phys. Rev. E\/} {\bf 99}(4) 042121
  \urlprefix\url{https://link.aps.org/doi/10.1103/PhysRevE.99.042121}

\bibitem{joulain2016quantum}
Joulain K, Drevillon J, Ezzahri Y and Ordonez-Miranda J 2016 {\em Phys. Rev.
  Lett.\/} {\bf 116}(20) 200601
  \urlprefix\url{https://link.aps.org/doi/10.1103/PhysRevLett.116.200601}

\bibitem{guo2018quantum}
Guo B~q, Liu T and Yu C~s 2018 {\em Phys. Rev. E\/} {\bf 98}(2) 022118
  \urlprefix\url{https://link.aps.org/doi/10.1103/PhysRevE.98.022118}

\bibitem{majland2020quantum}
Majland M, Christensen K~S and Zinner N~T 2020 {\em Phys. Rev. B\/} {\bf
  101}(18) 184510
  \urlprefix\url{https://link.aps.org/doi/10.1103/PhysRevB.101.184510}

\bibitem{PhysRevResearch.2.033285}
Naseem M~T, Misra A, M\"ustecaplio\ifmmode~\breve{g}\else \u{g}\fi{}lu O~E and
  Kurizki G 2020 {\em Phys. Rev. Research\/} {\bf 2}(3) 033285
  \urlprefix\url{https://link.aps.org/doi/10.1103/PhysRevResearch.2.033285}

\bibitem{wijesekara2020optically}
Wijesekara R~T, Gunapala S~D, Stockman M~I and Premaratne M 2020 {\em Phys.
  Rev. B\/} {\bf 101}(24) 245402
  \urlprefix\url{https://link.aps.org/doi/10.1103/PhysRevB.101.245402}

\bibitem{PhysRevA.103.052613}
Ghosh R, Ghoshal A and Sen U 2021 {\em Phys. Rev. A\/} {\bf 103}(5) 052613
  \urlprefix\url{https://link.aps.org/doi/10.1103/PhysRevA.103.052613}

\bibitem{wijesekara2021darlington}
Wijesekara R~T, Gunapala S~D and Premaratne M 2021 {\em Phys. Rev. B\/} {\bf
  104}(4) 045405
  \urlprefix\url{https://link.aps.org/doi/10.1103/PhysRevB.104.045405}

\bibitem{mandarino2021thermal}
Mandarino A, Joulain K, G\'omez M~D and Bellomo B 2021 {\em Phys. Rev.
  Applied\/} {\bf 16}(3) 034026
  \urlprefix\url{https://link.aps.org/doi/10.1103/PhysRevApplied.16.034026}

\bibitem{e24010032}
Liu Y~Q, Yu D~H and Yu C~S 2022 {\em Entropy\/} {\bf 24} ISSN 1099-4300
  \urlprefix\url{https://www.mdpi.com/1099-4300/24/1/32}

\bibitem{majer2007coupling}
Majer J, Chow J~M, Gambetta J~M, Koch J, Johnson B~R, Schreier J~A, Frunzio L,
  Schuster D~I, Houck A~A, Wallraff A, Blais A, Devoret M~H, Girvin S~M and
  Schoelkopf R~J 2007 {\em Nature\/} {\bf 449} 443--447
  \urlprefix\url{https://doi.org/10.1038/nature06184}

\bibitem{sillanpaa2007coherent}
Sillanp{\"a}{\"a} M~A, Park J~I and Simmonds R~W 2007 {\em Nature\/} {\bf 449}
  438--442 \urlprefix\url{https://doi.org/10.1038/nature06124}

\bibitem{you2011atomic}
You J~Q and Nori F 2011 {\em Nature\/} {\bf 474} 589--597
  \urlprefix\url{https://doi.org/10.1038/nature10122}

\bibitem{xiang2013hybrid}
Xiang Z~L, Ashhab S, You J~Q and Nori F 2013 {\em Rev. Mod. Phys.\/} {\bf
  85}(2) 623--653
  \urlprefix\url{https://link.aps.org/doi/10.1103/RevModPhys.85.623}

\bibitem{koski2014experimental}
Koski J~V, Maisi V~F, Pekola J~P and Averin D~V 2014 {\em Proc. Natl. Acad.
  Sci. U.S.A.\/} {\bf 111} 13786--13789
  \urlprefix\url{https://doi.org/10.1073/pnas.1406966111}

\bibitem{pekola2015towards}
Pekola J~P 2015 {\em Nat. Phys.\/} {\bf 11} 118--123
  \urlprefix\url{https://doi.org/10.1038/NPHYS3169}

\bibitem{hofer2016quantum}
Hofer P~P, Souquet J~R and Clerk A~A 2016 {\em Phys. Rev. B\/} {\bf 93}(4)
  041418 \urlprefix\url{https://link.aps.org/doi/10.1103/PhysRevB.93.041418}

\bibitem{kim2002entanglement}
Kim M~S, Lee J, Ahn D and Knight P~L 2002 {\em Phys. Rev. A\/} {\bf 65}(4)
  040101 \urlprefix\url{https://link.aps.org/doi/10.1103/PhysRevA.65.040101}

\bibitem{schaller2009transport}
Schaller G, Kie\ss{}lich G and Brandes T 2009 {\em Phys. Rev. B\/} {\bf 80}(24)
  245107 \urlprefix\url{https://link.aps.org/doi/10.1103/PhysRevB.80.245107}

\bibitem{liao2011quantum}
Liao J~Q, Huang J~F and Kuang L~M 2011 {\em Phys. Rev. A\/} {\bf 83}(5) 052110
  \urlprefix\url{https://link.aps.org/doi/10.1103/PhysRevA.83.052110}

\bibitem{2013Steady}
Bellomo B and Antezza M 2013 {\em Europhys. Lett.\/} {\bf 104} 236--247
  \urlprefix\url{https://iopscience.iop.org/article/10.1209/0295-5075/104/10006}

\bibitem{hewgill2018quantum}
Hewgill A, Ferraro A and De~Chiara G 2018 {\em Phys. Rev. A\/} {\bf 98}(4)
  042102 \urlprefix\url{https://link.aps.org/doi/10.1103/PhysRevA.98.042102}

\bibitem{man2019improving}
Man Z~X, Tavakoli A, Brask J~B and Xia Y~J 2019 {\em Phys. Scr.\/} {\bf 94}
  075101
  \urlprefix\url{https://iopscience.iop.org/article/10.1088/1402-4896/ab0c51/meta}

\bibitem{cattaneo2019local}
Cattaneo M, Giorgi G~L, Maniscalco S and Zambrini R 2019 {\em New J. Phys.\/}
  {\bf 21} 113045 \urlprefix\url{https://doi.org/10.1088/1367-2630/ab54ac}

\bibitem{dean2010boron}
Dean C~R, Young A~F, Meric I, Lee C, Wang L, Sorgenfrei S, Watanabe K,
  Taniguchi T, Kim P, Shepard K~L and Hone J 2010 {\em Nat. Nanotechnol.\/}
  {\bf 5} 722--726 \urlprefix\url{https://doi.org/10.1038/NNANO.2010.172}

\bibitem{li2014black}
Li L, Yu Y, Ye G~J, Ge Q, Ou X, Wu H, Feng D, Chen X~H and Zhang Y 2014 {\em
  Nat. Nanotechnol.\/} {\bf 9} 372--377
  \urlprefix\url{https://doi.org/10.1038/NNANO.2014.35}

\bibitem{zhang2017metallic}
Zhang Q, Li X, Ma Q, Zhang Q, Bai H, Yi W, Liu J, Han J and Xi G 2017 {\em Nat.
  Commun.\/} {\bf 8} 1--9 \urlprefix\url{https://doi.org/10.1038/ncomms14903}

\bibitem{geim2009graphene}
Geim A~K 2009 {\em Science\/} {\bf 324} 1530--1534
  \urlprefix\url{https://doi.org/10.1126/science.1158877}

\bibitem{faugeras2010thermal}
Faugeras C, Faugeras B, Orlita M, Potemski M, Nair R~R and Geim A~K 2010 {\em
  ACS Nano\/} {\bf 4} 1889--1892
  \urlprefix\url{https://doi.org/10.1021/nn9016229}

\bibitem{novoselov2012roadmap}
Novoselov K~S, Fal V~I, Colombo L, Gellert P~R, Schwab M~G and Kim K 2012 {\em
  Nature\/} {\bf 490} 192--200
  \urlprefix\url{https://doi.org/10.1038/nature11458}

\bibitem{efetov2018fast}
Efetov D~K, Shiue R~J, Gao Y, Skinner B, Walsh E~D, Choi H, Zheng J, Tan C,
  Grosso G, Peng C {\em et~al.\/} 2018 {\em Nat. Nanotechnol.\/} {\bf 13}
  797--801 \urlprefix\url{https://doi.org/10.1038/s41565-018-0169-0}

\bibitem{wang2020inverse}
Wang J, Casati G and Benenti G 2020 {\em Phys. Rev. Lett.\/} {\bf 124}(11)
  110607
  \urlprefix\url{https://link.aps.org/doi/10.1103/PhysRevLett.124.110607}

\bibitem{2003Local}
Laustsen T, Verstraete F and Enk S 2003 {\em Quantum Inf. Comput.\/} {\bf 3}
  64--83

\bibitem{gour2005deterministic}
Gour G, Meyer D~A and Sanders B~C 2005 {\em Phys. Rev. A\/} {\bf 72}(4) 042329
  \urlprefix\url{https://link.aps.org/doi/10.1103/PhysRevA.72.042329}

\bibitem{breuer2002theory}
Breuer H~P, Petruccione F {\em et~al.\/} 2002 {\em The theory of open quantum
  systems\/} (Oxford University Press on Demand)

\bibitem{li2012colloquium}
Li N, Ren J, Wang L, Zhang G, H\"anggi P and Li B 2012 {\em Rev. Mod. Phys.\/}
  {\bf 84}(3) 1045--1066
  \urlprefix\url{https://link.aps.org/doi/10.1103/RevModPhys.84.1045}

\bibitem{PhysRevE.95.022128}
Ordonez-Miranda J, Ezzahri Y and Joulain K 2017 {\em Phys. Rev. E\/} {\bf
  95}(2) 022128
  \urlprefix\url{https://link.aps.org/doi/10.1103/PhysRevE.95.022128}

\bibitem{PhysRevE.104.054137}
Upadhyay V, Naseem M~T, Marathe R and M\"ustecapl\ifmmode \imath \else \i
  \fi{}o\ifmmode~\breve{g}\else \u{g}\fi{}lu O~E 2021 {\em Phys. Rev. E\/} {\bf
  104}(5) 054137
  \urlprefix\url{https://link.aps.org/doi/10.1103/PhysRevE.104.054137}

\bibitem{manzano2019boosting}
Manzano G, Giorgi G~L, Fazio R and Zambrini R 2019 {\em New J. Phys.\/} {\bf
  21} 123026
  \urlprefix\url{https://iopscience.iop.org/article/10.1088/1367-2630/ab5c58}

\bibitem{PhysRevApplied.15.054050}
Iorio A, Strambini E, Haack G, Campisi M and Giazotto F 2021 {\em Phys. Rev.
  Applied\/} {\bf 15}(5) 054050
  \urlprefix\url{https://link.aps.org/doi/10.1103/PhysRevApplied.15.054050}

\bibitem{RevModPhys.77.633}
Fleischhauer M, Imamoglu A and Marangos J~P 2005 {\em Rev. Mod. Phys.\/} {\bf
  77}(2) 633--673
  \urlprefix\url{https://link.aps.org/doi/10.1103/RevModPhys.77.633}

\bibitem{nefedkin2022dark}
Nefedkin N, Cotrufo M, Krasnok A and Al{\`u} A 2022 {\em Adv. Quantum
  Technol.\/} {\bf 5} 2100112
  \urlprefix\url{https://onlinelibrary.wiley.com/doi/full/10.1002/qute.202100112}

\bibitem{PhysRevE.96.052126}
He Z~c, Huang X~y and Yu C~s 2017 {\em Phys. Rev. E\/} {\bf 96}(5) 052126
  \urlprefix\url{https://link.aps.org/doi/10.1103/PhysRevE.96.052126}

\bibitem{poulsen2022dark}
Poulsen K and Zinner N~T 2022 {\em arXiv preprint arXiv:2203.12623\/}
  \urlprefix\url{https://doi.org/10.48550/arXiv.2203.12623}

\end{thebibliography}

\end{document}